\documentclass[alpha-refs]{wiley-article} 

\usepackage{siunitx}
\usepackage[utf8]{inputenc} 
\usepackage[T1]{fontenc}    
\usepackage{hyperref}       
\usepackage{url}            
\usepackage{booktabs}       
\usepackage{amsfonts}       
\usepackage{nicefrac}       
\usepackage{microtype}      
\usepackage{lipsum}
\usepackage{graphicx}
\usepackage{caption}
\usepackage{subcaption}
\usepackage{amsmath} 

\usepackage{longtable} 

\usepackage{lscape} 
\usepackage{pdflscape} 

\usepackage{geometry}

\usepackage[table]{xcolor}

\usepackage{tikz}

\definecolor{gray(x11gray)}{rgb}{0.92, 0.92, 0.92} 

\papertype{Original Article} 
\paperfield{version 0.2}

\title{Bayesian item response models for citizen science ecological data.}

\author[1 2]{Edgar Santos-Fernandez, PhD}

\author[1 2]{Kerrie Mengersen, Dist Prof}

\affil[1]{School of Mathematical Sciences. Y Block, Floor 8, Gardens Point Campus.
		Queensland University of Technology. GPO Box 2434. Brisbane.
		QLD 4001. Australia.
		}
\affil[2]{Australian Research Council Centre of Excellence for Mathematical and
		Statistical Frontiers (ACEMS)}

\corraddress{School of Mathematical Sciences. Y Block, Floor 8, Gardens Point Campus.
		Queensland University of Technology. GPO Box 2434. Brisbane.
		QLD 4001. Australia.}
\corremail{santosfe@qut.edu.au}

\fundinginfo{
Australian Research Council (ARC) Laureate Fellowship Program ``Bayesian Learning for Decision Making in the Big Data Era'' (ID: FL150100150).

Australian Research Council Centre of Excellence for Mathematical and Statistical Frontiers (ACEMS)}

\runningauthor{Santos-Fernandez and Mengersen (2020)}

\begin{document}

\begin{frontmatter}
\maketitle

\begin{abstract}
\begin{enumerate}
\item So-called ``citizen science'' data elicited from crowds has become increasingly popular in many fields including ecology. However, the quality of this information is being frequently debated by many within the scientific community. Modern citizen science implementations therefore require measures of the users' proficiency that account
for the difficulty of the tasks. 
\item We introduce a new methodological framework of item response and linear logistic test models 
with application to citizen science data used in ecological research. This approach accommodates spatial autocorrelation within the item difficulties, and produces relevant ecological measures of species and site-related difficulties, discriminatory power and guessing behaviour. 
These, along with estimates of the subjects' abilities, allow better management of these programs and provide deeper insights.  
This paper also highlights the fit of item response models to big data via divide-and-conquer.
\item We found that the suggested methods outperform the traditional item response models in terms of RMSE, 
accuracy, and WAIC based on leave-one-out cross-validation on simulated and empirical data.
\item We present a comprehensive implementation using a case study of species identification in the Serengeti, Tanzania. The R and Stan codes are provided for full reproducibility.  
Multiple statistical illustrations and visualisations are given which allow extrapolation to a wide range of citizen science ecological problems.
\end{enumerate}
\end{abstract}
\end{frontmatter}

\section {Introduction}
\label{sec:Int}

Citizen science is becoming extremely valuable in modern science helping to
overcome some of the limitations in conventional science settings \citep{raykar2010learning, howe2008crowdsourcing, bonney2014next}. 
For example, large worldwide networks of participants are contributing to conservation efforts, generating large volumes of valuable information and achieving a better understanding and awareness of biodiversity \citep{mckinley2017citizen}. 
The numerous examples found in the literature range from the worldwide estimation of the abundance of birds \citep{sullivan2009ebird}, to species of mammals in Africa \citep{swanson2015snapshot},
the distribution of jaguars in the Peruvian Amazon \citep{mengersen2017modelling} and 
hard coral cover estimation in the Great Barrier Reef \citep{peterson2018monitoring}.
Plenty of projects can be found in online platforms e.g. iNaturalist, eButterfly, Zooniverse.

However, the quality of the data produced by volunteers is often questioned in the scientific community
\citep{bonney2014next, kosmala2018integrating}.
Contributors' commitment, abilities, training and effort along with the difficulty of the task affect their performance \citep{kelling2015can, dennis2017efficient}.
In crowdsourcing projects, there is a growing interest in assessing how well users can perform estimation and classification tasks and
specifically in measuring their latent abilities \citep[e.g.][]{falk2019evaluating}.

This can be approached using Item Response Theory (IRT) models,
which are special cases of the family of generalized mixed models and 
consist of the estimation of the latent participant's performance while accounting for the task's difficulty. 
There is an extensive literature on the use of IRT in multiple fields ranging from psychology, political and social sciences \citep{gadermann2012estimating, laurens2012psychotic, mcgann2014estimating}, and education to statistics and computer sciences \citep{wauters2010adaptive, de2010reducing, van2010irt, meyer2013fair}.  

Multiple frequentist and Bayesian approaches to IRT have been employed, along with a wide range of computational methods.
For example, for classification tasks, common models found in the literature are the logistic and normal models, while graded response models are employed for ordered categorical variables.  
Several implementations of Bayesian IRT models using different statistical languages have been also published.
For instance,  \citet{curtis2010bugs} used BUGS/JAGS, 
\citet{grant2016fitting, luo2018using} and \citet{burkner2019bayesian} used Stan while
 \citet{stone2015bayesian} employed SAS. 
Multiple examples can be found in \citet{fox2010bayesian} and in the Stan user manual \citep{carpenter2017stan}.

Latent variable models that have been used in ecology to describe the occurrence of species as a function of latent predictors \citep{pollock2014understanding, warton2015so} are related to the IRT models. The use, however, different formulations. 
However, not much attention has been paid to estimating the abilities and performance measures for citizen science in ecological settings. 

Measures of participants' abilities and their differentiation are relevant for several reasons. 
Statistical inference from crowdsourced data assigns greater weights to information from more competent users
\citep{kosmala2016assessing, bird2014statistical, johnston2018estimates}.
In the absence of a gold standard, ability-derived weights can be used in consensus voting algorithms or
for building reputation and leaderboard systems \citep{silvertown2015crowdsourcing, callaghan2019improving}.
Extremely low abilities are generally associated with careless, non-genuine respondents and bots which are generally excluded for being non-informative and representing a burden for inference and computation.

In this research, subjects are referred to as participants, citizens, annotators, users, while
items refer to information sources (images, videos, audio files, etc.) 
\citep{zhang2013managing, gura2013citizen, ratnieks2016data}. 
However, we concentrate on the particular case species identification on images.

Data collected in citizen science projects is challenging compared to what is found in the traditional item response literature. For example, the number of items for classifications can be large, reaching often hundreds of thousands and 
        in some cases, it could be larger than the number of users.
        Additionally, participants score a different number of items. Some users engage heavily while others contribute lightly \citep{hsing2018economical}. However, a user rarely scores all the items if the number is large.
        As a consequence, generally, a large proportion of data is missing. 
        Overall, the skills of the respondents are generally heterogeneous, ranging from experts to beginners.

These complications are exacerbated when the focus of citizen science is on the classification of features or the identification of species in images.
In this setup, camera, site, and species-specific factors can affect the underlying item difficulty level \citep{willi2019identifying}. 
Several image specific components maybe also critical e.g.: the camera resolution, brand, flash effect, etc. 
Other crucial factors are the visibility, time of the day (day or night), landscape and the vegetation.
It has been also well documented that some species are far more difficult to identify than others. 
Look-alike or closely related species are more challenging, leading frequently to misclassifications \citep{chambert2018two, hsing2018economical}.

In ecological research, items such as images are usually georeferenced \citep{mckinley2017citizen, callaghan2019improving}.
An added complication in such a setting is that items that are geographically close to each other tend to have more similar characteristics such as the probability of containing the target species, compared to those that are further away. 
For this reason, many citizen science applications account for spatial autocorrelation. Some examples can be found in \citet{pagel2014quantifying, purse2015landscape, arab2015spatio, humphreys2019seasonal, altwegg2019occupancy}.
Latent image difficulties tend to exhibit spatial auto-correlation, which should be accounted for in the IRT models. 
Ignoring such autocorrelation can lead to biased parameter estimates 
and underestimating the associated standard errors and hence potentially erroneous inference.
See e.g. \citet{lichstein2002spatial, ver2018spatial}.

The identification of areas with spatial correlation is also relevant for survey administration purposes. 
Distinguishing areas in which the items have relatively high difficulties is vital to provide recommendations on how to improve the quality of the images, generate more useful training and users' documentation, 
and to better manage the order in which tasks are presented for classification, etc.
Images from camera traps difficult to identify might indicate technical issues. 
Additionally, identification of images with no visible species and covered by vegetation is relevant for data cleaning purposes.

\subsubsection* {\bf A brief literature review of IRT models}  

Since the introduction of the Rasch model \citep{rasch1960studies}, several extensions and variations have been proposed for the IRT. 
The three parameter logistic model (3PL) \citep{birnbaum1968some} and the graded response model by \cite{samejima1969estimation}
are among the most relevant contributions. 
Many examples and implementations can be found in widely known published textbooks e.g. \citet{baker2004item, van2013handbook, embretson2013item, Ayalade2013theory}.
IRT modelling has been approached from the generalized linear model (GLM) perspective and a very didactic discussion can be found in \citet{wilson2008explanatory}.
In a spatial context, \citet{canccado2016item} has suggested a model that borrows some principles of spatial statistics for the identification of spatial clusters. 
Another model accounting for changes in space and time is presented in \citet{juhl2019measurement}. 

A new class of state-space model called dynamic item response is introduced by \citet{wang2013bayesian} 
for longitudinal data. Under the principle that the abilities change over time, 
this dynamic approach produces a growth curve for the latent trait. 
Another dynamic IRT model was presented by \citet{weng2018real},  
where model parameters are updated as the data becomes available in a sequential manner.  

Within the Bayesian philosophy, multiple model implementations have been suggested e.g. \citet{patz1999applications, fox2010bayesian, albert2015introduction}.
Possibly the greatest limitation of Bayesian item response models is the identifiability issues
arising from having a large number of parameters to be estimated from the data.
For a detailed discussion on identifiability see e.g. \citet{fox2010bayesian}.
Strategies like anchoring the sum or the mean of the difficulties to zero have been suggested to cope with this limitation.

In this research, we introduce extensions of the 3PL item response model (U)sing (S)patially dependent item difficulties (henceforth 3PLUS) and variations of the linear logistic test model (LLTM).
These approaches are tailored to citizen science data elicited from georeferenced images.
We discuss the advantages and limitations of the suggested methods.
We take a Bayesian approach for modelling using prior information for the parameters of interest. 
This has multiple advantages which includes better parameter estimates for cases when the number of samples for
some users is relatively small \citep{mislevy1986bayes, choy2009elicitation}. 
We start with a simulation study to compare the proposed model to the state of the art. 
This is followed by a case study of citizen science image classification in the Serengeti, Africa.

\section{Methods} 
In the case of citizen-elicited classification in an image, a common approach is to select points 
and ask the user whether each point contains the target class or species. 
Let the binary response variable $Y$ represents whether the question was correctly answered 
by the $i ^\textrm{th}$ citizen ($i = 1,\cdots,I$), in the $j^\textrm{th}$ image ($j = 1,\cdots,J$) for the $k^\textrm{th}$ point ($k = 1,\cdots,K$), $Y_{ijk} \sim \textrm{Bern} \left(p_{ijk} \right)$.

In some cases the interest is on identifying species in the whole image and the above case is reduced to $k=1$.
The probability $p_{ijk}$ of a correct answer can be modeled using the three-parameter logistic model \citep{lord2012applications, baker2004item}. 

\begin{equation}
p_{ijk} = \eta_j + \left(1 - \eta_j \right)\frac{1}{1+\textrm{exp}\left \{ -\alpha_j\left ( \theta_i - b_j\right ) \right \}}.
\label {eq:eq11}
\end{equation}

Here $\theta_i$ is the latent ability of the $i^{\textrm{th}}$ user, such that 
the higher the $\theta_i$ values are, the larger the probability of identifying correctly the class of the item.
The parameter $b_j$ represents the difficulty of the $j^{\textrm{th}}$ image and 
$\alpha_j > 0$ represents the slope or discrimination parameter of image $j$ 
indicating how quickly the function will go from 0 to 1.
The parameter $\eta_j$ is the {\it pseudoguessing} parameter of the $j^{\textrm{th}}$ image that is the lower asymptote, which
accounts for users' chance of answering correctly by guessing. 

\begin{figure*}[htbp]
	\centering
		\includegraphics[width=5in]{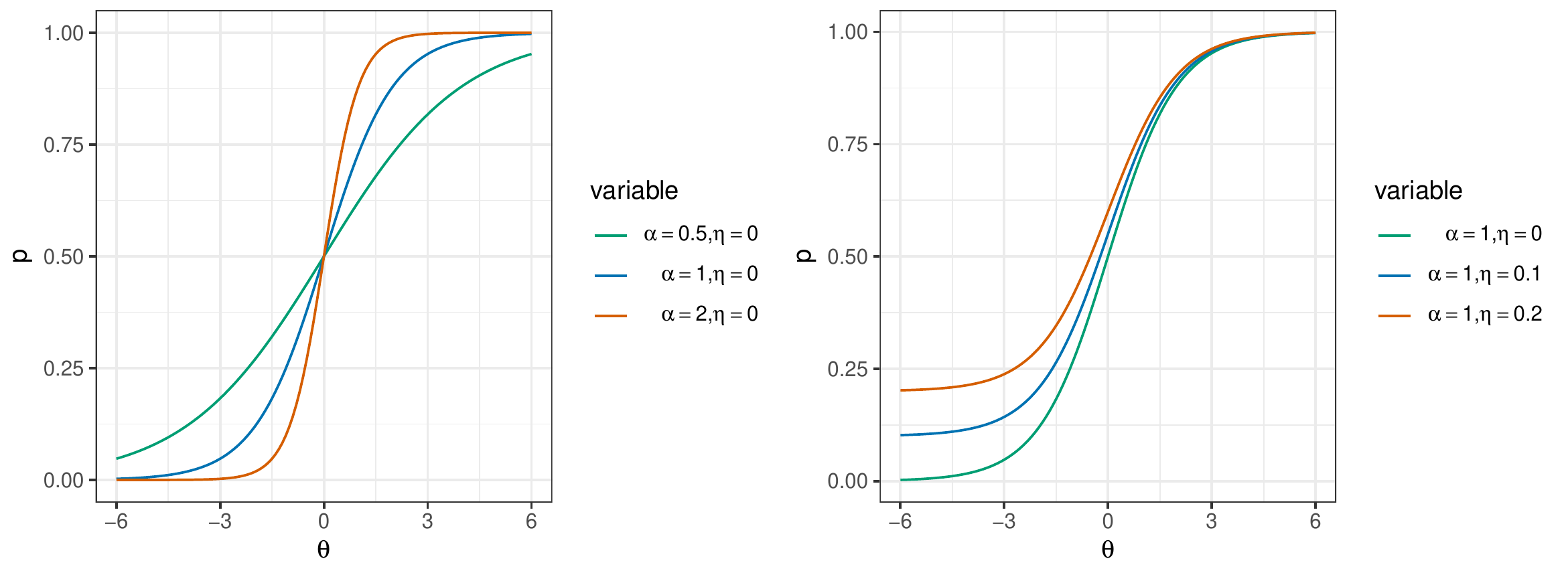}
	\caption{Item characteristic curves of the 3PL model giving the probability of correctly answering a standard item ($b=0$) as a function of the latent user ability $\theta$. In (a) we compare curves from three different values of the slope $\alpha$ with no guessing i.e. $\eta = 0$. 
	In (b) we compare curves using three pseudoguessing values $\eta = 0, 0.1, 0.2$ and a fixed value of $\alpha = 1$.  } %
	\label{fig:Fig_curves}
\end{figure*}

Several variations of this model can be found in the literature \citep{baker2004item, Ayalade2013theory}. For example, 
the two-parameter logistic (2PL) model is a reduction arising from setting $\eta_j = 0$, 
while the Rasch model is obtained using $\eta_j = 0$ and $\alpha_j = 1$ \citep{Cai2016}. 
Another popular approach replaces the logistic link function by the probit link \citep{albert1992bayesian}. 
The difficulty, slope and pseudoguessing parameter are considered locally homogenous within the image, so,
the parameters $b_j$, $\alpha_j$ and $\eta_j$ are common for each point $k$.   
The effect of these parameters on $p_{ijk}$ is shown in Fig.\ref{fig:Fig_curves}. 

In this paper, we suggest a spatial extension of the 3PL model in which we model the difficulties as a 
random spatial effect.  We adopt a conditional autoregressive (CAR) prior \citep{besag1991},
noting that the general approach can accommodate other representations of the spatial autocorrelation.
CAR priors have been found to adequately capture spatial variability in citizen science applications \citep{pagel2014quantifying, purse2015landscape,arab2016spatio, arab2015spatio, altwegg2019occupancy}.
However, multiple other alternatives such as Gaussian random fields \citep{humphreys2019seasonal} and stochastic partial differential equations (SPDE) \citep{peterson2020monitoring} can be found in the literature.
Under the CAR prior, the distribution of $b$ 
conditional on the first-order neighbours is the average of the $n_l$ first-order neighbours plus Gaussian noise.



\begin{equation} 
b_{l}|b_{m},\tau_b \sim \mathcal N \left ( \frac{1}{n_l}\sum_{l \sim m}^{.}b_m, \frac{1}{\tau_b n_l} \right )
\label {eq:eq3001}
\end{equation}
 
\noindent where $n_l$ is the number of neighbours for region $l$, $l \sim m$ means $l$ and $m$ are neighbours and $l \neq m$. The parameter $\tau_b$ is the precision of the spatial effect.

\subsection{LLTM model extensions}
We propose a further extension of the 3PLUS model called item explanatory, which is a special case of
 the linear logistic test model (LLTM) \citep{fischer1973linear, wilson2008explanatory}. 
In this approach, the item parameter ($b_{j}$) is a linear combination of factors governing 
the difficulty ($\sum_{m=1}^{M}\beta_mX_{jm}$),
where $M$ is the number of factors. 
This model does not give directly an indication of the difficulties associated with the images, 
but explains the effect of the item related factors.
For instance, often we are interested in the camera difficulty that accounts for the image quality and on the complications associated with the site (unique latitude and longitude) in which the camera is placed rather than on the individual image difficulties.

Similarly, the intrinsic difficulty associated with the species can be considered.
This explains the misclassification error due to species that look like the target one (e.g. Grant's vs Thomson's gazelle) and also mimic animal-specific behavior and features.

The probability that citizen $i$ correctly identifies species $l$ in the $j^{\textrm{th}}$ site is given by:

\begin{equation} \label{eq:pijl}
p_{ijl} = \eta_. + \left(1 - \eta_. \right)\frac{1}{1+\textrm{exp}\left \{ -\alpha_.\left ( \theta_i - \beta_jI_{j} - \beta_lI_{l}\right ) \right \}}
\end{equation}

\noindent where $I_j$ and $I_l$ are indicator variables of the site and species respectively. 
The parameters $\beta_j$ and $\beta_l$ are difficulties associated with the site and species respectively.
The pseudoguessing $\eta_.$ can be associated with the species or with the sites. 
That is why we indexed it with a dot (.). 
The first case is more reasonable since it explains what species are more likely to be correctly classified due to guessing. 
High values of species-related pseudoguessing values indicate species that are easily classified for those users with less proficiency.
Similarly, species or site-specific discrimination parameters can be estimated in the model ($\alpha_l$ or $\alpha_j$),
showing which items are more suitable for differentiating the users.

To avoid confusion, we will retain the notation previously used to refer to the models but we add a prime symbol (')
to denote \emph{test model} extensions.   
That is, the test model (3PLUS') of Eq.\ref{eq:pijl} is an extension of the model introduced in 
Eq.\ref{eq:eq11} with $b_j$ replaced by $\beta_jI_{j} + \beta_lI_{l}$.

\section{Simulation study}

We first present a simulation study to compare the performance of the 3PL and 3PLUS models. 
Consider $j = 225$ unique geographical locations in a 15 $\times$ 15 grid, which is generally large enough for the illustration and has been employed in several practical designs e.g. \citep{swanson2015snapshot}.
For the image difficulties we generated the spatial autocorrelation using a multivariate normal distribution, $b_j \sim \textrm{MVN},\left( \mu_b = 0, \Sigma_b \right)$, with a spatial covariance matrix $\Sigma_b$ obtained using the simple exponential form e.g.: $\Sigma_b = \exp(-d/r)$, where $d$ is the Euclidean distance between locations and $r$ is a scaling parameter.

Fig.\ref{fig:diff_voronoi} shows Voronoi polygons using the image locations as centroids and boundaries based on Euclidean distance \citep{okabe2009spatial, gold2016spatial}, which were used to define the spatial domain in the model.
In (a) we show the spatial association between the difficulties and 
regions characterized by clusters of images that are {\it easy} (light blue) and {\it hard} (dark blue) to classify.
Consider five groups of users with fixed abilities: 
$\theta_{.} = \left\{-1, -0.5,0 , 0.5, 1 \right\}$. 
We set several values for the discrimination ($\alpha$) ranging from small to large slopes, $\alpha_{.} = \left\{0.25, 0.50, \cdots, 1.50, 1.75 \right\}$. 
We assume having six target classes and consider that users will have on average 1 in 6 chances of guessing the true target class
with extremely low knowledge. Therefore, a pseudoguessing parameter $\eta \sim \textrm{Beta}\left(1,5\right)$ is employed, which is a 
weakly informative prior commonly used in the literature for this parameter.

A random user, slope and pseudoguessing value is assigned with replacement to each image id.  
Each citizen scores multiple images and each image is classified several times.
Every image contains 15 elicitation points. Using $p$ from Eq.\ref{eq:eq11} we simulated binary realizations using the Bernoulli distribution.

\subsection{Fitting the Bayesian models}

We want to learn about the latent parameters $\theta_i$, $b_j$, $\alpha_j$ and $\eta_j$ using the observed variable ($Y_{ijk}$).
We employ Markov chain Monte Carlo simulations, in particular Hamiltonian Monte Carlo (HMC), in 
the software package Stan \citep{carpenter2017stan} which is based on the no-U-turn sampler (NUTS) \citep{hoffman2014no}.
We used 3 chains each of 50,000 samples after we discarded a burn-in of 20,000 samples.
We used the prior distributions shown in Fig \ref{fig:priors}. 
The CAR prior was implemented according to \citet{morris2019bayesian}. 

\begin{figure*}[h]
	\centering
	{\scriptsize
\begin{align*}
		\theta_{i}&\sim \mathcal{N}\left(0,\sigma_{\theta}\right)   && \text{\# hierarchical informative prior on the abilities}\\
		\sigma_{\theta}&\sim \textrm{uniform}\left(0,10\right)  && \text{\# flat prior for the user's abilities sd }\\
		b_{3PL}&\sim \mathcal{N}\left(\mu_{b},\sigma_{b}\right)  && \text{\# informative prior on the difficulties in 3PL model}\\	
    \mu_{b}&\sim \mathcal{N}\left(0,5\right) && \text{\# normal prior for the mean of the item difficulty }\\
		\sigma_{b}&\sim \textrm{Cauchy}\left(0,5\right)T(0,)  && \text{\# flat prior for the sd of the item difficulty }\\
		b_{\textrm{3PLUS}} &\sim \textrm{CAR}\left(\tau, W, D\right)  && \text{\# CAR prior for the spatial model}\\		 
		\tau_{b}  & \sim \textrm{Gamma}\left(1, 1\right) && \text{\# precision of the spatial effect } \\	
		\alpha &\sim \mathcal{N}\left(1,\sigma_{\alpha}\right) && \text{\# normal prior with mean 1 on the slope}\\
		\sigma_{\alpha}&\sim \textrm{Cauchy}\left(0,5\right)T(0,)  && \text{\# half Cauchy prior for the slope sd, where $T(0,)$ means lower truncation at 0 }\\
		\eta & \sim \textrm{beta}\left(1,5\right)   &&  \text{\# weakly informative prior on the pseudoguessing}\\
	  \tau &  && \text{\# precision parameter in the CAR prior}\\ 
     D & _{m \times m} &&  \text{\# diagonal matrix}\\
     W & _{m \times m} &&  \text{\# adjacency matrix}\\ 
		\label{eq:66}
\end{align*} 
}%
	\caption{Prior distributions used in the models. } %
	\label{fig:priors}
\end{figure*}

The parameter $b_{3PL}$ is the difficulty in the traditional model (3PL), while $b_{\textrm{3PLUS}}$ represents the 
difficulty in the spatial approach.
We anchored the ability of the ``reference'' user by setting it equal to zero. 
This was done by fitting the model in the \emph{mirt} package \citep{mirt} and finding the user with the score closest to zero.

The comparison of the models was based on the following criteria: 
(1) confusion matrix and the accuracy when estimating the parameters (difficulties, abilities, slopes, pseudoguessing), 
(2) Root Mean Square Error (RMSE), (3) Watanabe–Akaike information criterion (WAIC) \citep{watanabe2010asymptotic}.

\subsection{Simulation results}

\begin{figure}[htb]
\begin{subfigure}{1\textwidth}
\centering
\caption{Boxplot of the difficulties }  
   \includegraphics[width=4.7in]{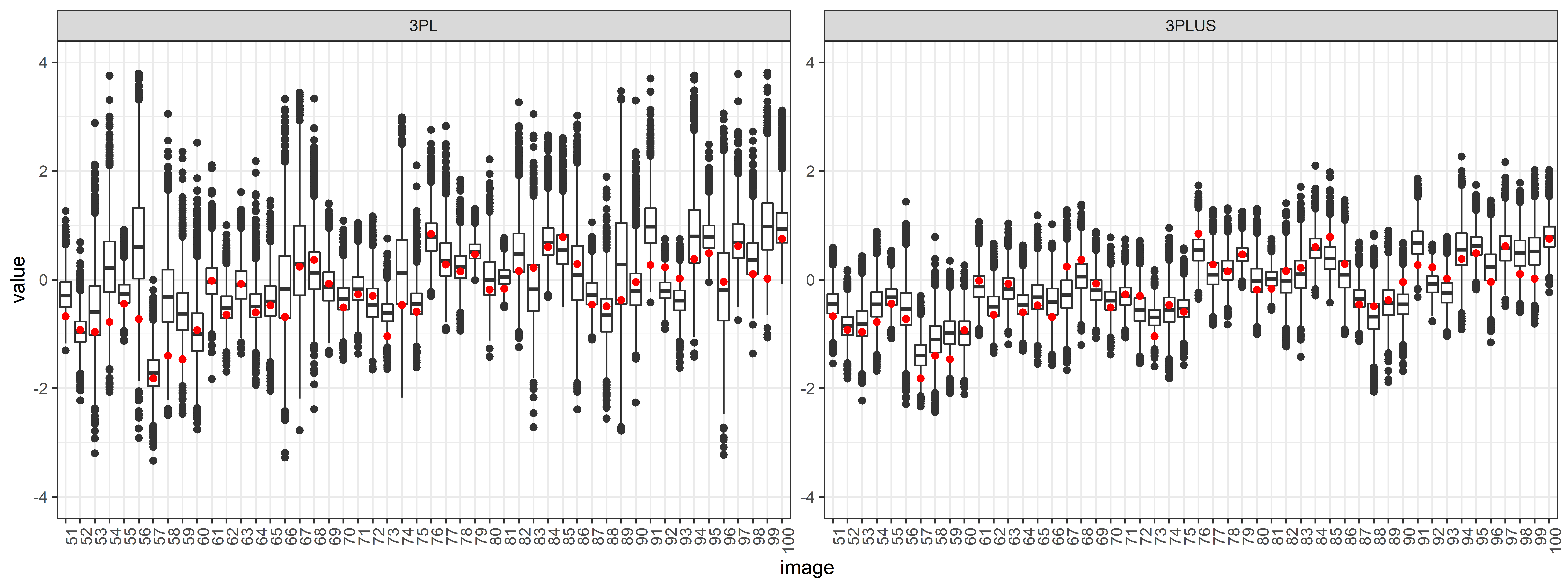}
   \label{fig:Fig_diff_boxplot} 
\end{subfigure}
\begin{subfigure}{.45\textwidth}
  \centering
	\caption{Slope}
  \includegraphics[width=0.8\linewidth]{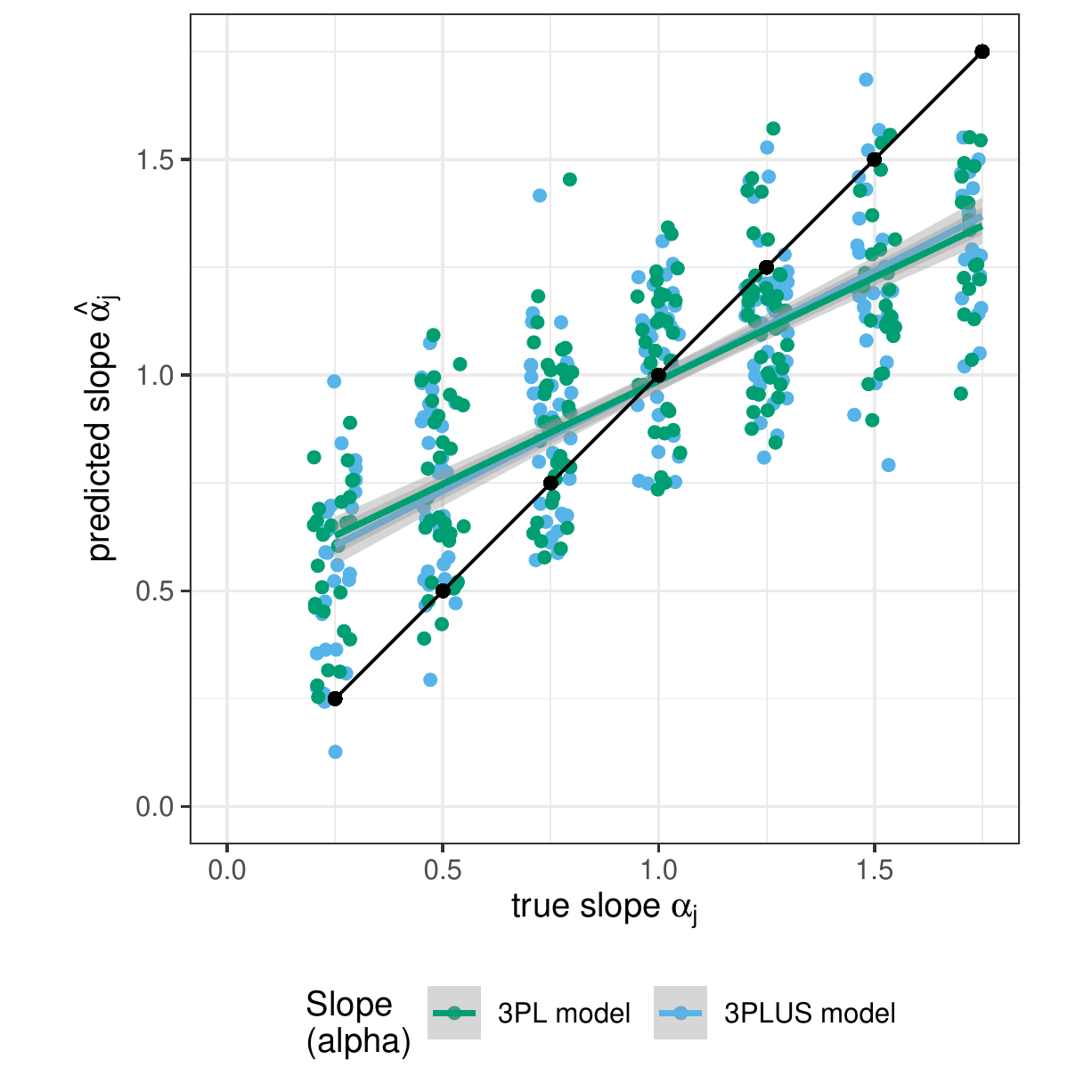}
  \label{fig:Fig_alpha}
\end{subfigure}%
\begin{subfigure}{.45\textwidth}
  \centering
\caption{Guessing} 
 \includegraphics[width=0.8\linewidth]{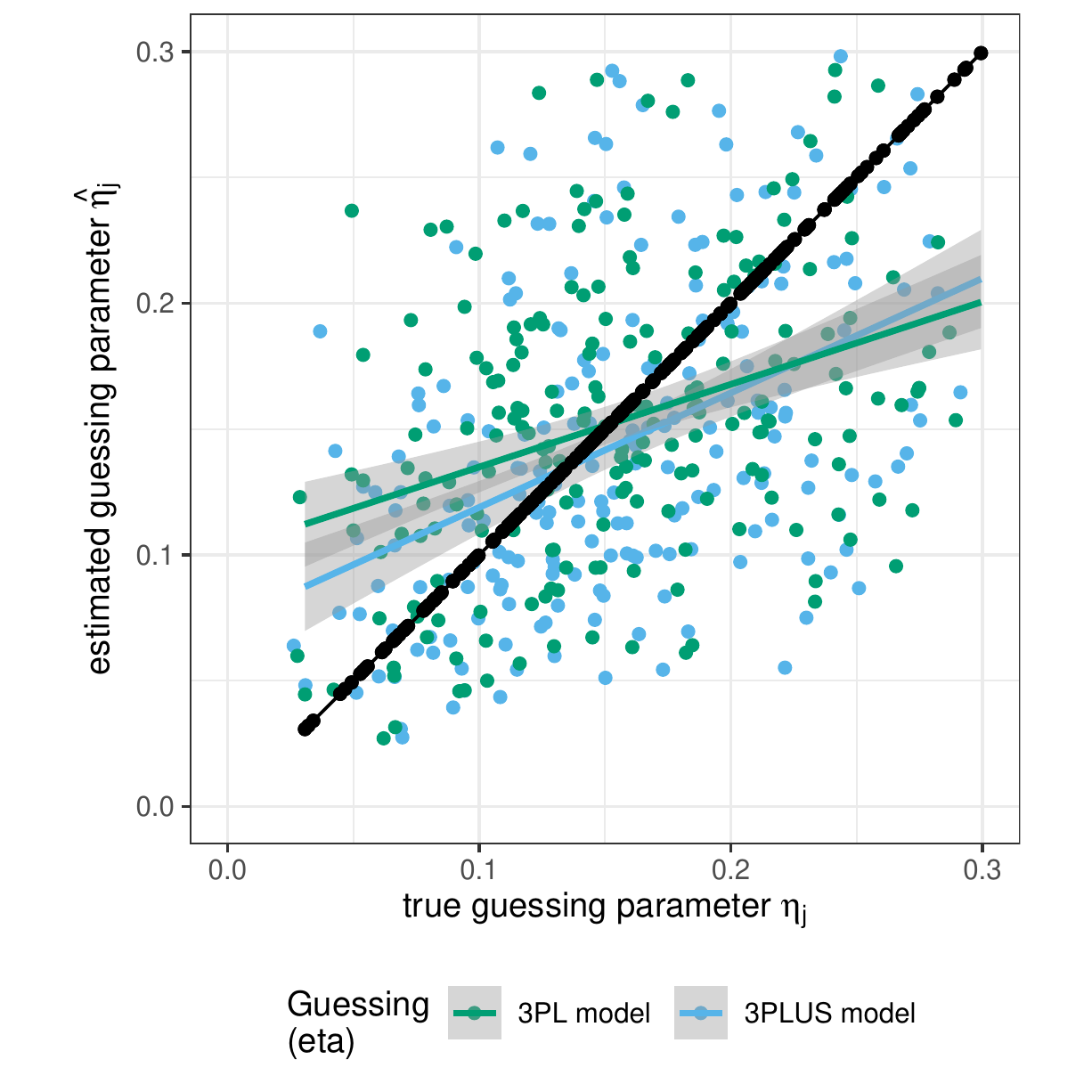}
  \label{fig:Fig_eta}
\end{subfigure}
\caption{{\scriptsize(a) Boxplot of the estimated difficulties for locations 51-100. The true parameter value (red dots) in the traditional 3PL (left) and the 3PLUS model (right).Estimated slope (b) and guessing (c) parameters in the traditional 3PL (green) and the 3PLUS model (blue).
The y-axis gives the estimates while the x-axis represents the true value. The ideal case of perfect estimation is represented in black color.}
}
\end{figure}

Both models produced similar user abilities estimates (See Fig \ref{fig:abil_violin_3PL} and \ref{fig:abil_violin_spat} in the Supplementary Material).
The RMSE for the 3PLUS model was slightly smaller  (0.2036 vs 0.2092).  	
The 3PLUS model produces better difficulty estimates (Fig \ref{fig:Fig_diff_boxplot} and \ref{fig:diff_voronoi} ) with a better accuracy and precision. 
	The yellow labels in Fig \ref{fig:diff_voronoi} (b) and (c) are the locations where the true difficulties estimates were not correctly identified.  
	The 3PLUS model achieved an 80\% prediction accuracy of the difficulty classes compared to 62.22\% under the 3PL model.
	This model also halved the RMSE: $\textrm{RMSE}_{\textrm{3PLUS}} = $ 0.2596 vs. $\textrm{RMSE}_{\textrm{3PL}} = 0.4808$. 
	The correlations between the estimated difficulties and the true values were $r^{2}_{\textrm{3PLUS}} = 0.9685$ vs 
	$r^{2}_{\textrm{3PL}} = 0.8776$.
	Both models produced similar estimates of the slope ($\alpha$), but the pseudoguessing parameter ($\eta$)
 was better estimated in the 3PLUS model in terms of RMSE. See Fig \ref{fig:Fig_alpha} and \ref{fig:Fig_eta}. 
The 3PLUS model also produces a slightly smaller WAIC (44,586.7 vs 44,637.0) and LOO (leave-one-out cross-validation) values (44,588.1 vs 44,638.7).

\subsection{Effect of having more classifications}

One of the main strengths of citizen science programs is the rapid acquisition of a large volume of classifications \citep{bonney2009citizen}.
However, several issues relevant to these programs are not often quantified. 
For instance, how does the number of participants affect the precision of the estimates?
How many sampled locations are needed?
What is a suitable number of classifications per image? 
We designed an experiment to assess the effects of these three factors on the model fit and considered the following values:
   
\begin{enumerate}
\item grid size: This represents the number of unique geo-tagged locations for the items, e.g. image GPS coordinates: 10$\times$10, 15$\times$15 and 25$\times$25.
\item number of participants: 9, 21 and 60. This was chosen to be very small, small and medium based on the impact or the precision of the estimates. 
\item the number of elicitation points per image: This frequently arises in approaches used in ecology and is also known as the random point count methodology \citep[e.g.][]{kohler2006coral} : 5, 15 and 25. 
\end{enumerate}

\begin{figure*}[htbp]
	\centering
		\includegraphics[width=6.0in]{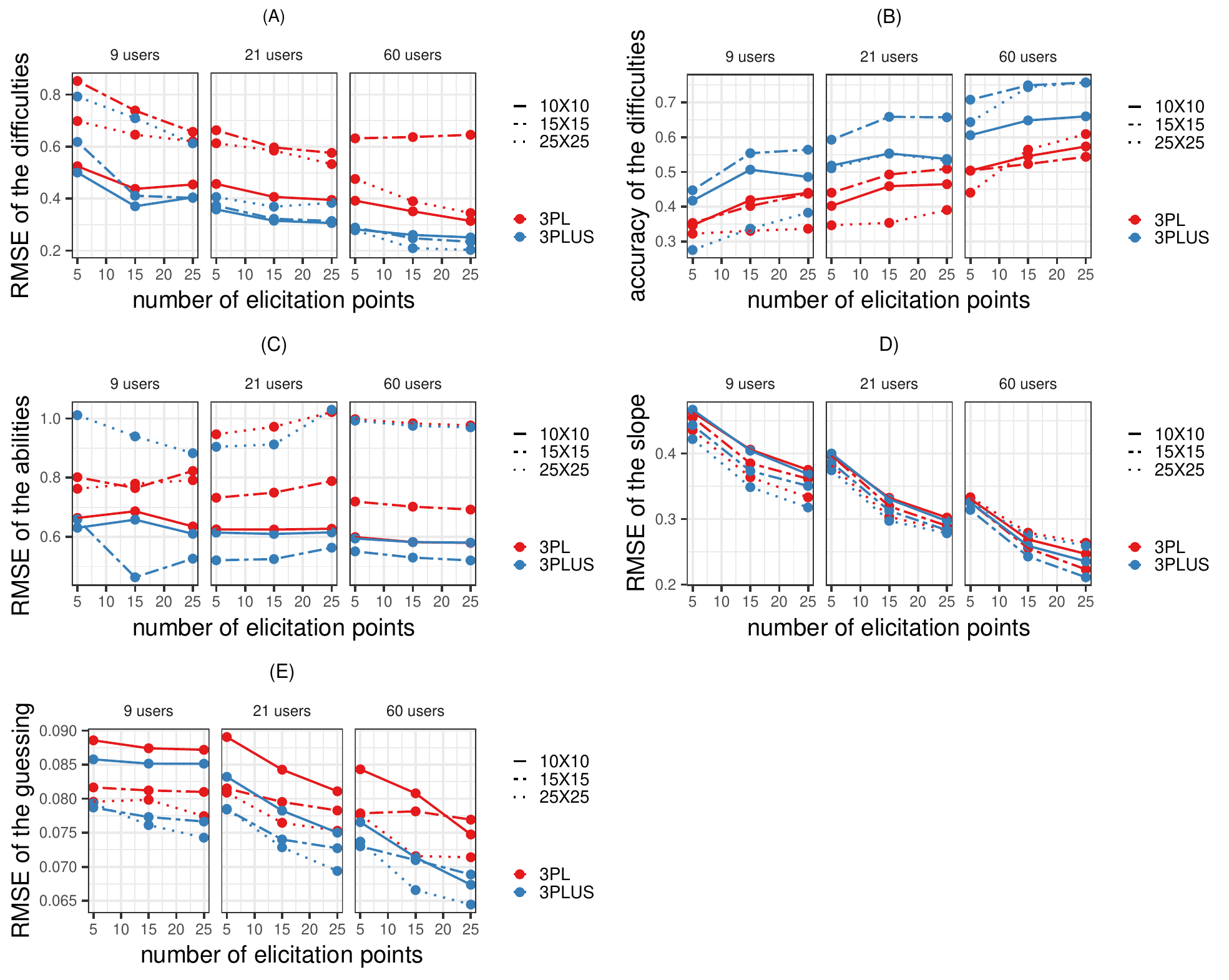}
	\caption{Root mean square error (RMSE) of the difficulties (A), abilities (C), slopes (D) and pseudoguessing (E) and difficulties accuracy (B) for each model (3PL vs 3PLUS). 
The x-axis represents the number of elicitation points/image and the panels give the number of users on each group. 
The grid size is plotted with different line types.
} 
	\label{fig:exp_all}
\end{figure*}

For this combination of parameters, we generated 27 datasets and fit both the 3PL and the 3PLUS models.
Fig \ref{fig:exp_all} depicts a comparison between the two models based on 
(a) RMSE of the difficulties and 
(b) the accuracy of retrieving the true category.
The three grid sizes are represented with different line types.
The panels show the number of users.
The proposed model produces a substantially smaller RMSE and better accuracy for most of the factor combinations.
The greater the number of elicitations and users the smaller the RMSE and the larger the accuracy is in both models.  

We fitted Bayesian regression models for the parameters (difficulties, abilities, slopes, and pseudoguessing) to the (log) RMSE
 as a function of the factors plus the model type using as a 
baseline factor the benchmark (3PL).
The posterior density and trace-plots showed well-mixed chains and convergence.
In the model for the log RMSE of the difficulties (Table \ref{table:model_results}), there are more differences between the models than when accounting for the factors.
The numbers of elicitations, citizens and the grid size were found to affect the RMSE. 
We found similar outcomes for the models using as a response variable the RMSE in the abilities, slopes, and pseudoguessing.
A negative value in the model dummy variable means that a reduced RMSE is obtained when using the 3PLUS model.
Larger numbers of elicitations and citizens also decrease the RMSE.

Fig \ref{fig:exp_all} compares the RMSE of the difficulties, abilities, slopes and pseudoguessing parameters and 
the accuracy retrieving the difficulties in both models.
The 3PLUS model produces smaller RMSE and the difference tends to be proportional to the number of users and the grid size. 

New algorithms are often computationally intensive and involve larger processing times. 
We performed the simulations on a High-Performance Computing (HPC) system with CPUs architecture Intel AVX and AVX-2.
No substantial differences in computing times were found between models.

\section{ \emph{Hakuna my data}: A case study of species identification in the Serengeti}

\subsection{The Serengeti data}

\citet{swanson2015snapshot} describe a citizen science project that identified species from the Serengeti, Tanzania.  
This project is hosted on Zooniverse (\url{https://www.zooniverse.org/projects/zooniverse/snapshot-serengeti}) and captured more than a million images, with a total of 10.8 million classifications produced by 28,000 users.
Images were obtained by camera traps in 225 spatial locations. 
The data contains image details of the classifications per user and includes the spatial component given by the location of the cameras. 
\citet{swanson2015snapshot} published several datasets including the original raw classification data, the consensus voting results and a gold standard dataset. 
The gold standard dataset included classifications by experts of 4,140 images.
The proportion of correct classification per site showed spatial autocorrelation based on the Moran's I test ($p < 0.001$).

\begin{figure*}[h]
	\centering
		\includegraphics[width=5in]{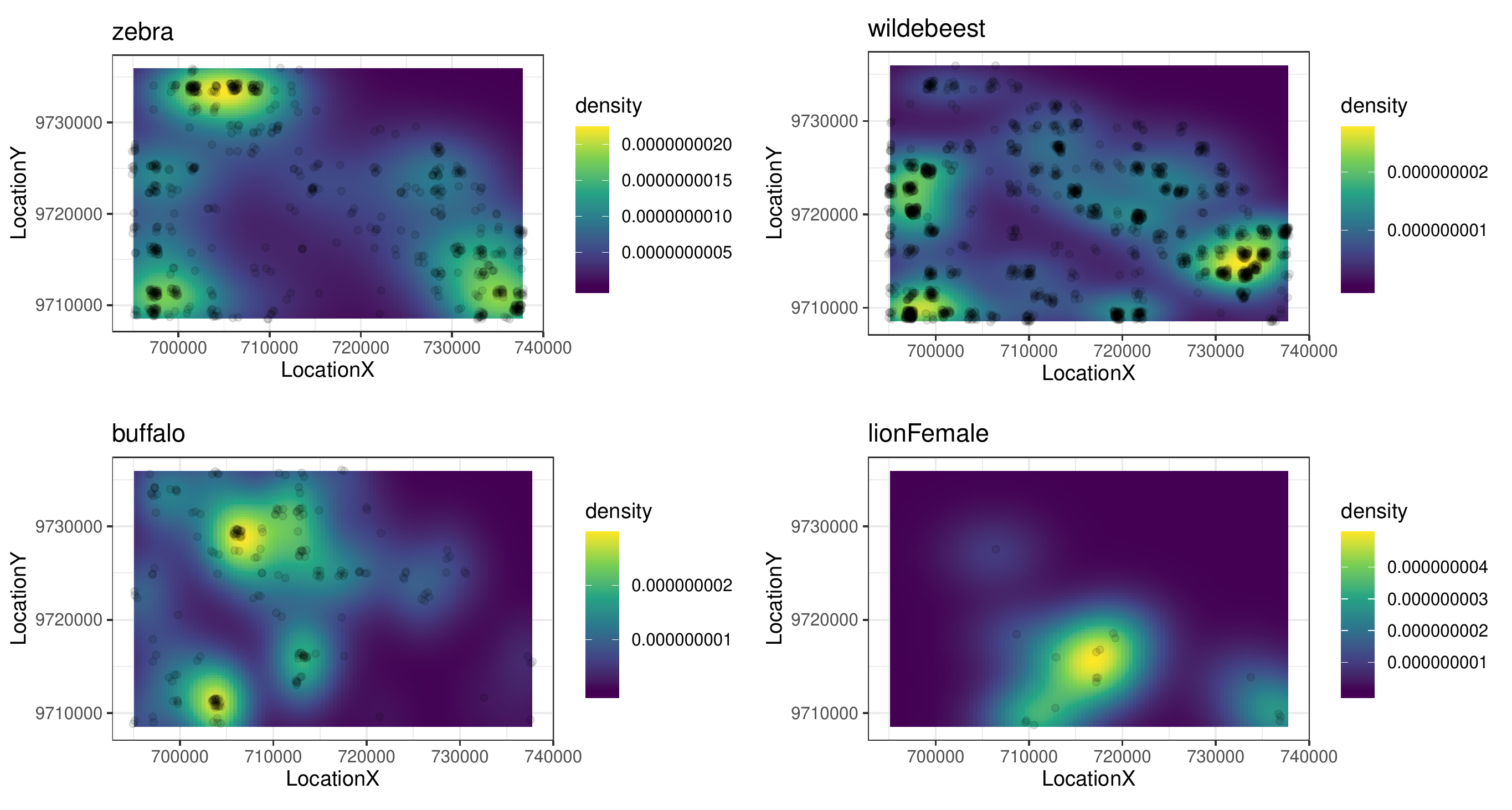}
	\caption{2D kernel density plot of the three most abundant species (prey) in the gold standard dataset plus female lion (predators). } 
	\label{fig:2d_density_plot_species}
\end{figure*}

A total of 50 species categories were identified including wildebeest, zebra, buffalo, hartebeest, lion (female and male), etc. 
The categories \emph{impossible} to identify and \emph{human} were also included.
The authors noted that subjects tend to use more the \emph{nothing here} classification rather than guessing when they were not sure of the species. 
Two-dimensional kernel density plots of the three most abundant species (prey) plus female lion (predators) is shown in Fig \ref{fig:2d_density_plot_species}.

Images were captured at different times of the day, including night time. 
Several other factors affecting the difficulty of the images are included: animal moving or feeding, the presence of babies, 
the relative position regarding the vegetation and camera placement, etc.

We selected the registered users with more than five classifications so that we can obtain suitable estimates of the abilities. 
For inclusion, images had to have a known location or site id.
This produced a dataset composed of almost 5 million observations, from 21,347 citizens from 225 locations.
The data and codes associated with the case study can be found in the repository \url{https://github.com/EdgarSantos-Fernandez/staircase}.

\subsection{Analysis using the gold standard data and model comparison}
Assessing the fit of several model variations and comparisons becomes difficult with large datasets. 
We first find the best performing model using the gold standard dataset and then fit it using the whole data. 
For simplicity, images with more than one species were excluded resulting in 88,517 observations. 

We considered 10 model variations of Eq.\ref{eq:pijl}, shown in Table \ref{tab:mods}.
We started with the 3PL and 3PLUS previously compared in the simulations.
Eight linear logistic test models were studied, from a very simple (1PL') to more complex (3PLUS').
Three variations of the 3PLUS' are examined, resulting from permutations of the site and species slopes and pseudoguessings parameters.
For instance, 3PLUS-2' and 3PLUS-3' involves a species-based guessing ($\eta_l$), while 
the slopes are site specific for 3PLUS-2' ($\alpha_{j}$) and species dependent for 3PLUS-3' ($\alpha_{l}$).

The 10 competing models were fit in Stan using four chains, 4000 iterations and a burn-in of 2000 reaching convergence.
The estimates for the effective number of parameters $\hat{p}_{waic}$ were found to be unreliable, exceeding 0.4 \citep{vehtari2017practical}, for which using leave-one-out cross-validation (LOO) is recommended.
The LOO measure also seems appropriate since we are interested in the applicability to other users/items. 
See Table \ref{tab:mods}.

According to the LOO information criterion values, the 3PLUS-2' model seems to be superior.
In this model, the pseudoguessing parameter is species-related,  while the slope is indexed to the site that accounts for the camera and location-specific factors.

\begin{table}[ht]
\centering
\caption{ {\scriptsize Model variations from the model in Eq.\ref{eq:pijl} ($p_{ijl}  = \eta_. + \left(1 - \eta_. \right)\frac{1}{1+\textrm{exp}\left \{ -\alpha_.\left ( \theta_i - \beta_jI_{j} - \beta_lI_{l}\right ) \right \}}$) where 
the citizen is indexed by $i$, the site location by $j$ and the species by $l$.
The symbol ' indicates test model extension.
} }
\scalebox{0.70}{
\begin{tabular}{lllrr}
  \hline
 Model & Name & Parameters & WAIC & LOO\\   \hline
 3PL & traditional 3PL model & $\eta_l$, $\alpha_l$, $\theta_i$, $\beta_l$, $\beta_j = 0$ & 94,396.2 & 94,443.9 \\  
 3PLUS &  spatial 3PL model  & $\eta_j$, $\alpha_j$, $\theta_i$, $\beta_j \sim \textrm{CAR}(W,D,\tau)$ and $\beta_l = 0$ & 86,153.3& 86,205.8\\ 

 1PL' & traditional 1PL test model &  $\eta = 0$, $\alpha = 1$, $\theta_i$, $\beta_l$ and $\beta_j$ & 86,167.6 & 86,197.2  \\  
 2PL' & traditional 2PL test model  & $\eta = 0$, $\alpha_j$, $\theta_i$, $\beta_l$ and $\beta_j$ & 85,401.2 & 85,453.6 \\ 
 3PL' & traditional 3PL test model&  $\eta_j$, $\alpha_j$, $\theta_i$, $\beta_l$ and $\beta_j$ & 85,439.2 & 85,495.7 \\
 
 1PLUS' & spatial 1PL test model & $\eta = 0$, $\alpha = 1$, $\theta_i$, $\beta_l$ and $\beta_j \sim \textrm{CAR}(W,D,\tau)$ & 86,186.0 & 86,215.8  \\ 
 2PLUS' & spatial 2PL test model &  $\eta = 0$, $\alpha_j$, $\theta_i$, $\beta_l$ and $\beta_j \sim \textrm{CAR}(W,D,\tau)$  & 85,411.2 & 85,463.3 \\   
 3PLUS-1' &  spatial 3PL test model  & $\eta_j$, $\alpha_j$, $\theta_i$, $\beta_l$ and $\beta_j \sim \textrm{CAR}(W,D,\tau)$ & 85,419.6 & 85,474.8 \\   
 3PLUS-2' &  spatial 3PL test model &  $\eta_l$, $\alpha_j$, $\theta_i$, $\beta_l$ and $\beta_j \sim \textrm{CAR}(W,D,\tau)$ & {\bf 85,199.3} & {\bf 85,257.9} \\   
 3PLUS-3' &  spatial 3PL test model &  $\eta_l$, $\alpha_l$, $\theta_i$, $\beta_l$ and $\beta_j \sim \textrm{CAR}(W,D,\tau)$ &  85,711.2 & 85,766.9 \\  
	\hline
\label{tab:mods}
	\end{tabular}	}
\end{table}

\subsection{Item response modeling for big data}
Fitting item response models to big datasets can be challenging and prohibitive within the Bayesian paradigm.
Operations involving a large number of parameters often exceed the memory, processing and disk capacities, 
which is exacerbated if simulation-based computation/estimation such as MCMC is used.
This is even more problematic when accounting for spatial autocorrelation \citep{katzfuss2012bayesian, datta2016hierarchical, finley2017applying}.
  
In this section, we focus on modelling datasets too large
to be fit directly in one machine or even on a modern HPC node. 
We use the dataset previously described, which is composed of almost 5 million classifications.  
The true answer (species) was obtained in the non-gold standard images using the majority vote.  

We took a divide-and-conquer approach, also known as divide-and-recombine  \citep{scott2016bayes, neiswanger2013asymptotically, wang2016statistical} splitting the big dataset into multiple shards or subsets. 
Models were fit to the independent subsets on independent machines with no communication, and subposterior estimates were then combined into global estimates using consensus Monte Carlo \citep{scott2016bayes}. 
Specifically, we used weighted averages of the posterior MCMC chains, with weights inversely proportional to the variance of the posterior samples. 
At each iteration, the consensus estimate of a parameter is given by $\theta_g = \left(\sum_{s}^{S}W_s \right)^{-1} \sum_{s}^{S}W_s\theta_{sg}$, where $s$ is the subset or machine, $g$ is the iteration number, 
$\theta_{sg}$ is the estimate at the iteration $g$ from the machine $s$ and 
$W_s = \textrm{Var}\left(\theta|y_s\right)^{-1}$.
The consensus Monte Carlo is known to produce an asymptotic approximation to the posterior based on the whole dataset \citep{scott2016bayes}. 
However, multiple other options could have been explored to obtain the combined subposterior distributions  \citep[e.g.][]{neiswanger2013asymptotically}.

The number of observations of each location and species is large. 
However, the classifications per user varied from a few to several hundred. 
Thus, we employed stratified sampling with users as grouping criteria thereby splitting the dataset into 10 equal parts. 
This yielded 10 shards with approximately half a million observations per shard. 
Some users will not be represented on all the subsets, while most of the species and sites were estimated on every shard.

All the results from this section can be reproduced using the  R Markdown file in the repository \url{https://github.com/EdgarSantos-Fernandez/staircase}, which illustrate (1) how to request resources on an HPC, (2) how to fit the models in Stan and (3) how to combine the posterior estimates. 

Shards were fitted in parallel with no communication on a high performance computing (HPC) system using 10 CPUs, three processors and a request of 280 Gb of memory.
Estimates of the latent parameters were obtained using the best performing model (3PLUS-2') found in the previous section.
We used three chains, a warm-up period of 3000 samples and 8000 iterations.
It took approximately 42 hours to fit each model and to compute the WAIC.
For combining the subposteriors into the whole dataset posterior, we modified the function \emph{consensusMCindep()} from the R package \emph{parallelMCMCcombine} \citep{parallelMCMCcombine}.
This approach is suitable for normally distributed parameters (difficulties, abilities, slopes) and
 has been found to work satisfactorily also for non-normal variables \citep{scott2016bayes} such as the pseudoguessing which is modelled using beta distributions. 
Summary statistics of some of the parameters of interest are shown in Table \ref{table:posterior_con}.

\begin{figure*}[h]
	\centering
		\includegraphics[width=3in]{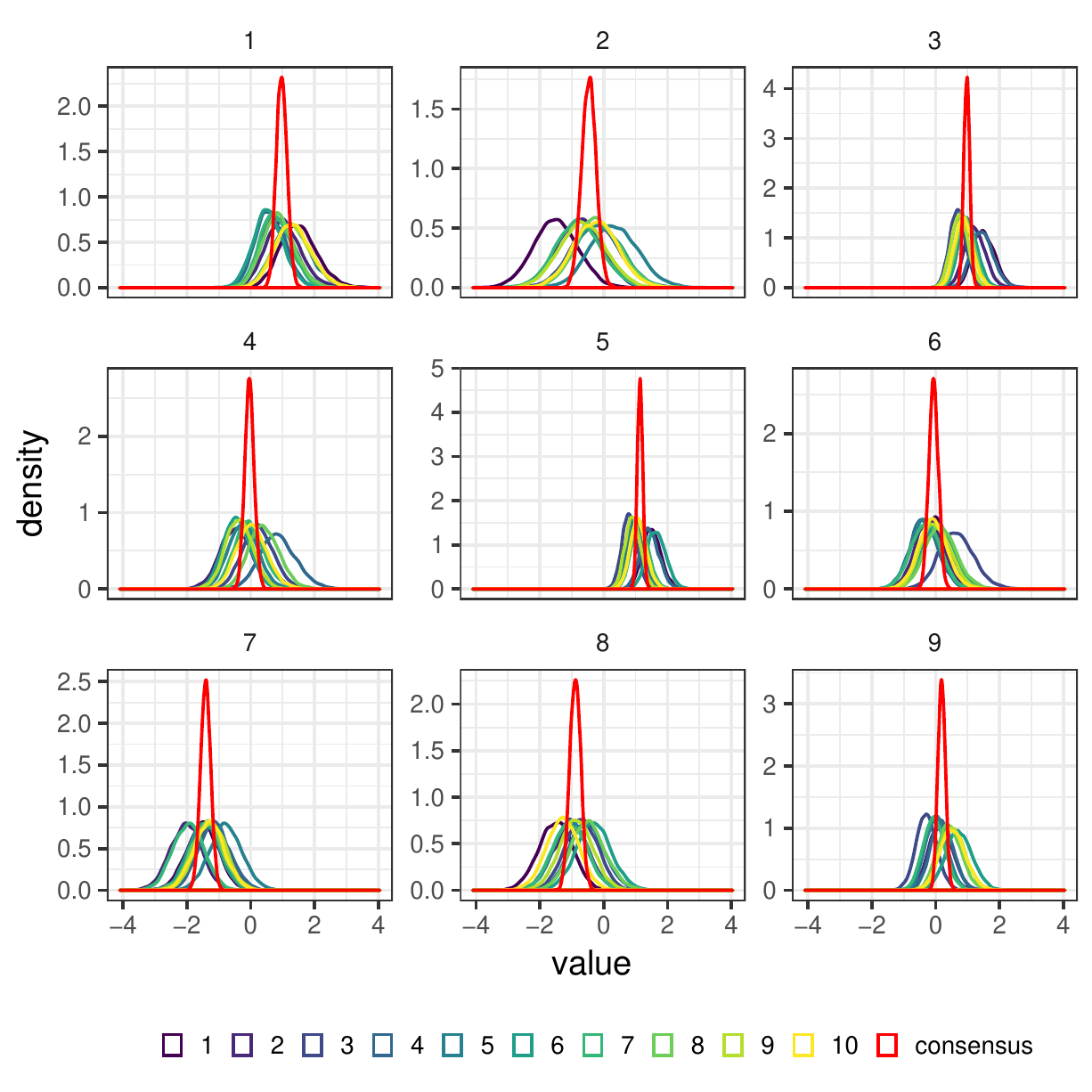}
	\caption{Posterior estimates of the abilities for nine users in 10 machines and the posterior obtained using 
	the consensus approach (in red).} 
	\label{fig:abil_densities_posterior}
\end{figure*}

On each machine, we obtained subposterior estimates for each of the parameters.
For instance, in Fig \ref{fig:abil_densities_posterior} we show the estimates for nine users with good, moderate and poor classification skills (three on each group).
In some cases, there could be a moderate variability on the estimates obtained from different shards e.g. user = 2.
This is because by random, users will have a better performance in some subsets than in the others.
See Fig \ref{fig:abil_ggpairs} for scatter plots comparing the posterior ability estimates in the 10 shards (abil1, abil2, etc.) and
in the recombined consensus posterior (abil).

\begin{figure*}[h]
	\centering
		\includegraphics[width=3.1in]{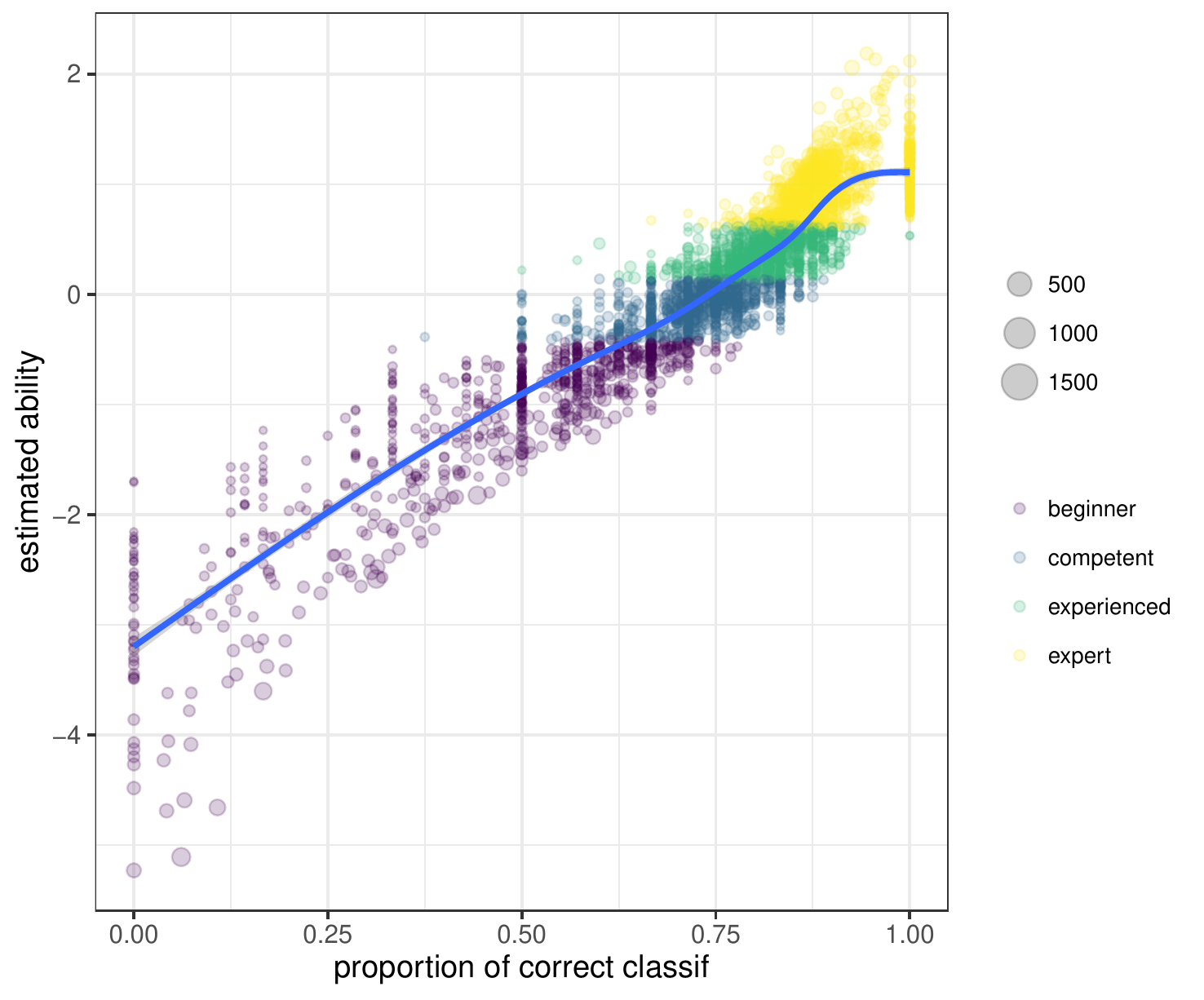}
	\caption{Posterior estimates of the users' abilities vs their proportion of correct classifications. 
	The size of the dots represents the users' number of classifications.} %
	\label{fig:seren_abil}
\end{figure*}

Fig \ref{fig:seren_abil} shows the user proficiency levels as a function of the proportion of correct classifications. 
Clustering the participants into groups based on their abilities is relevant for citizen science programs and crowdsourcing.
For instance, the latent classifications are often obtained by majority or consensus vote. 
Weighted variations of the voting systems can be easily implemented using as weights the posterior abilities with some penalization based on the variability of these estimates.
More effective compensation, rewards and bonuses can be readily implemented for example in projects using platforms such as Amazon Mechanical Turk. 
Finally, ability assessment is a powerful resource to detect software bots, careless participants, etc, which generally have the lowest ability scores (beginner group in Fig \ref{fig:seren_abil}).

\begin{figure}
\centering
\begin{subfigure}[b]{1\textwidth}
\centering
	 \caption{Subposterior estimates of the species difficulties}
		\includegraphics[width=4.5in]{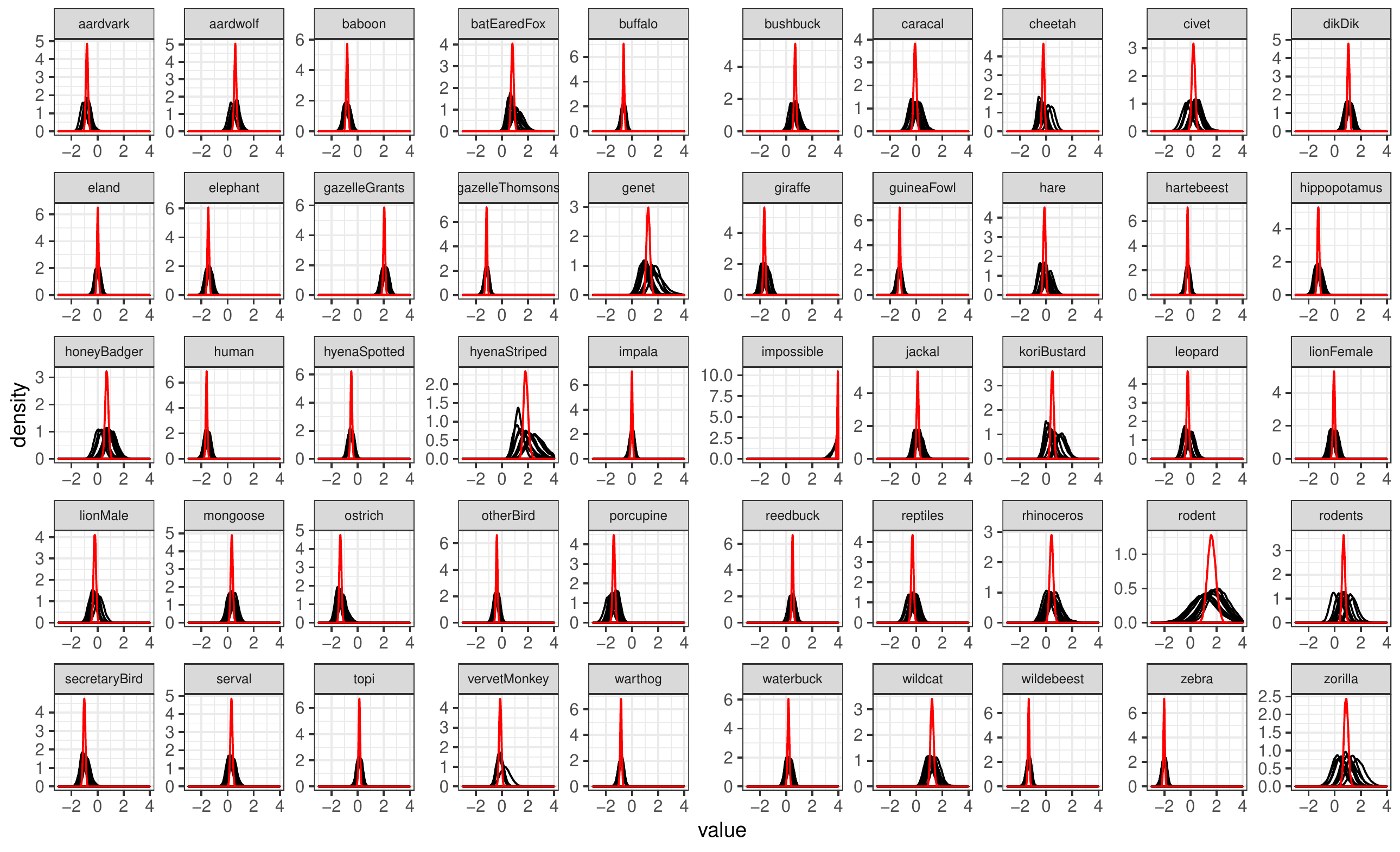}
     	\label{fig:species_densities_posterior_diff}
\end{subfigure}

\begin{subfigure}[b]{1\textwidth}
\caption{Species difficulties vs the proportion of correct classification}
\centering
   	\includegraphics[width=4.5in]{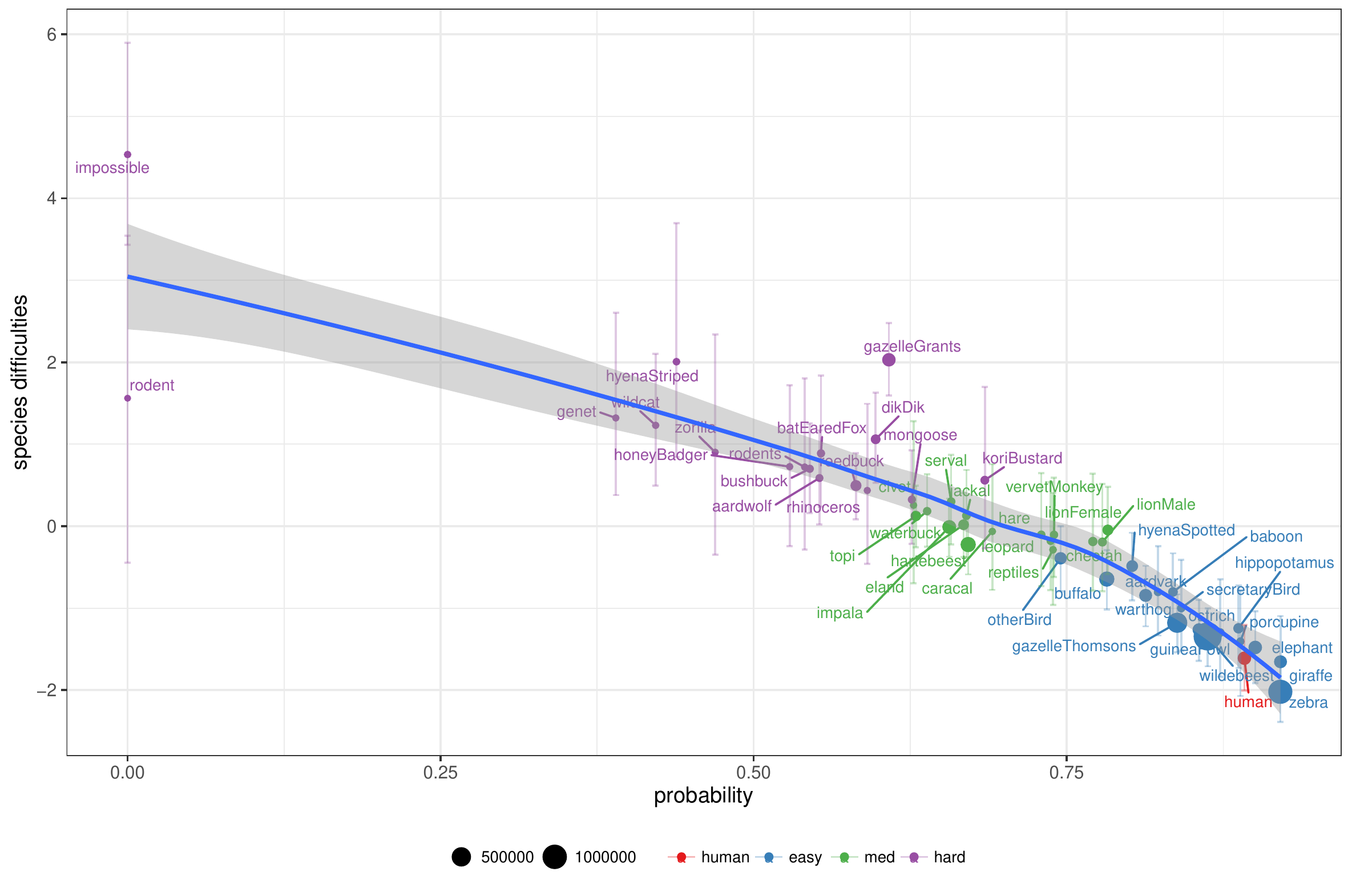}
   \label{fig:prob_vs_species}
\end{subfigure}

	\caption{ (a) Comparisons of the subposterior estimates of the species difficulties from the 10 subsets and from the consensus Monte Carlo (in red color) in the 50 species. 
	(b) Species difficulties and 95\% highest posterior density interval as a function of the proportion of species correct classification. 
	The size of the dots represents the number of times a given species is presented for classification.
	The category \emph{human} is used as a baseline.} 
	
\end{figure}

Similarly, we obtained posterior distributions for the difficulties in identifying the species (Fig \ref{fig:species_densities_posterior_diff}). Species with mean positive values are more difficult to identify e.g. hyena striped.
Other categories with negative estimates such as giraffe and zebra had a high probability of correct classification. 
Fig \ref{fig:prob_vs_species} shows the posterior estimates of the species difficulties as a function of the proportion of correct classification. On average citizens misclassified images containing humans with probability 0.10 and this could be considered 
as an indicator of careless responses.

\begin{figure}
\centering
\begin{subfigure}[b]{1\textwidth}
	 \caption{Subposterior estimates of the species pseudoguessing}
	\centering
		\includegraphics[width=4.5in]{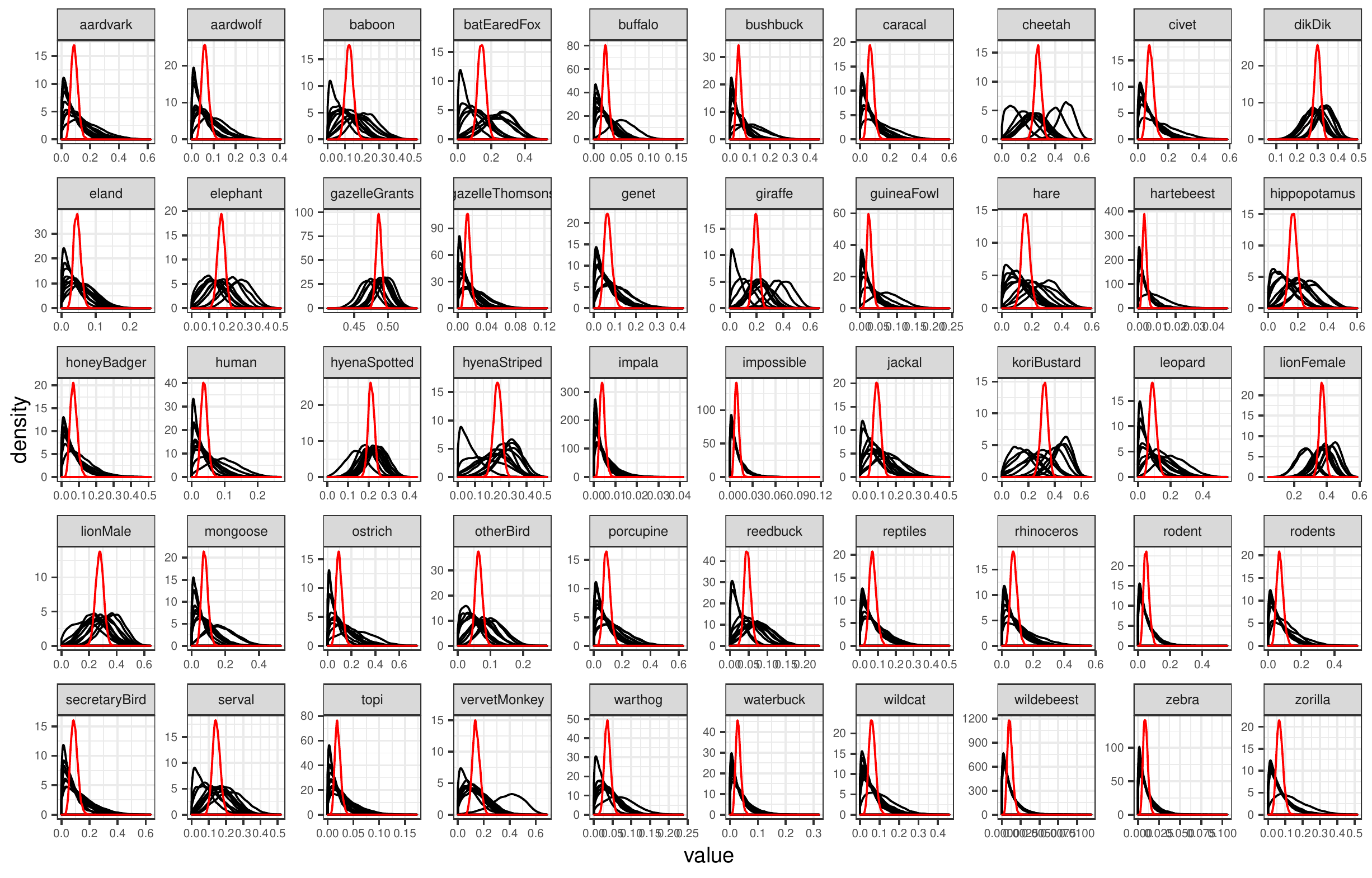}
     	\label{fig:guess_densities_posterior}
\end{subfigure}
\begin{subfigure}[b]{1\textwidth}
\caption{Species pseudoguessing vs the proportion of correct classification}
\centering
\includegraphics[width=5.0in]{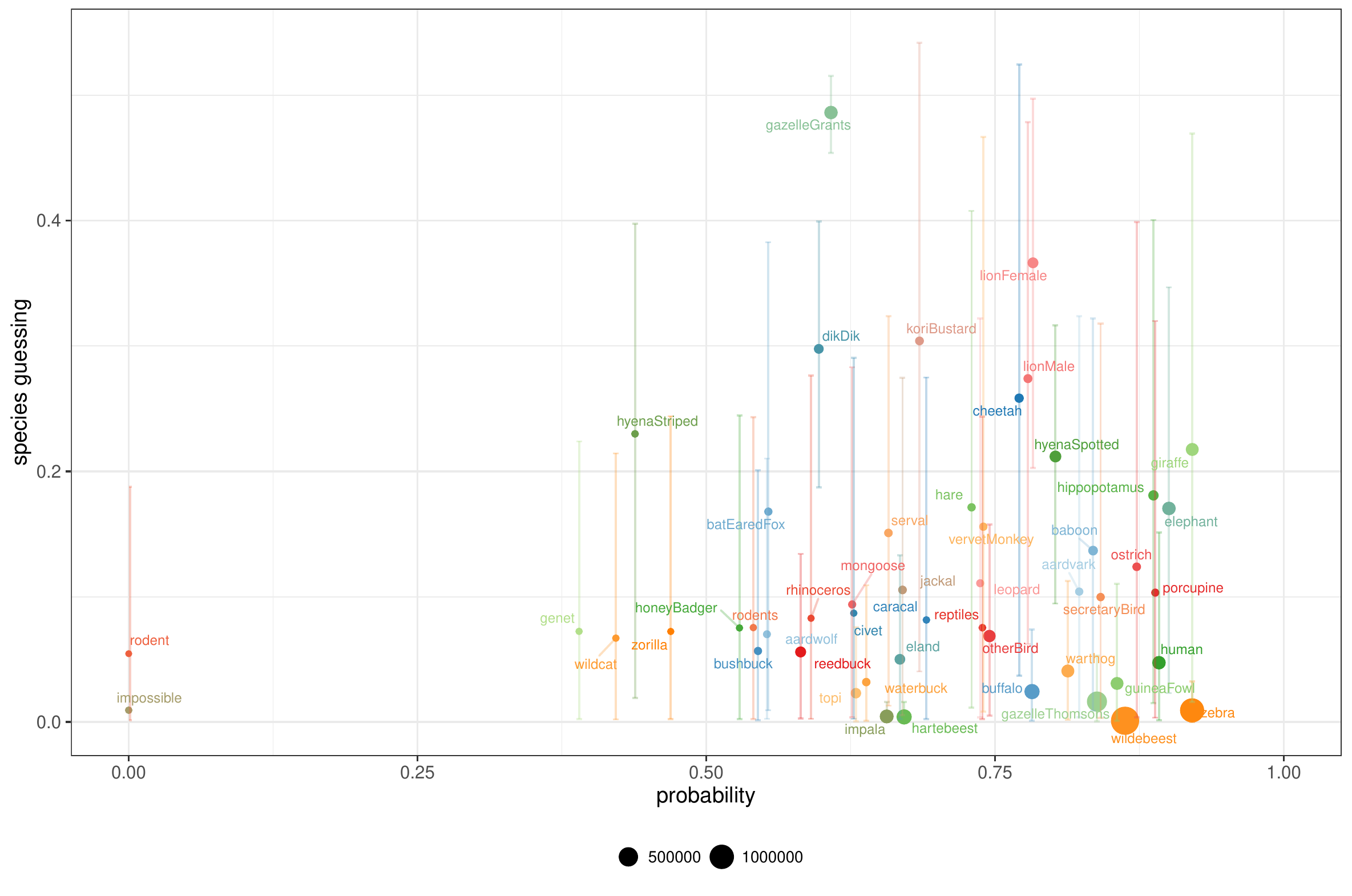}
   \label{fig:prob_vs_guess}
\end{subfigure}
	\caption{{\scriptsize (a) Comparisons of the posterior distributions of the species pseudoguessing in the subsets and compared to the combined estimate (in red color). (b) Posterior mean and highest density interval of the species pseudoguessing vs the proportion of correct classification. The size of the dots represent the number of times the species appeared in the images for classification.} }
	\end{figure}

We also obtained the minimum probability of a correct answer for species due to guessing, for users with
 extremely low abilities (Fig.\ref{fig:guess_densities_posterior} and \ref{fig:prob_vs_guess}).
Interestingly, large pseudoguessing estimates are obtained for lions, 
the only species for which the gender was required (female and male). 
These species had a higher probability of being correctly identified by less experienced users.
Fig \ref{fig:guess_species_ggpairs} displays scatter plots of the correlation between shards, 
showing a high correlation between the combined estimates (guess\_all) and the ones obtained from the shards (guess1, guess2, etc.). 


\begin{figure}
\centering
\begin{subfigure}[b]{1\textwidth}
\centering
	 \caption{Subposterior estimates of the site difficulties}
	\includegraphics[width=4.25in]{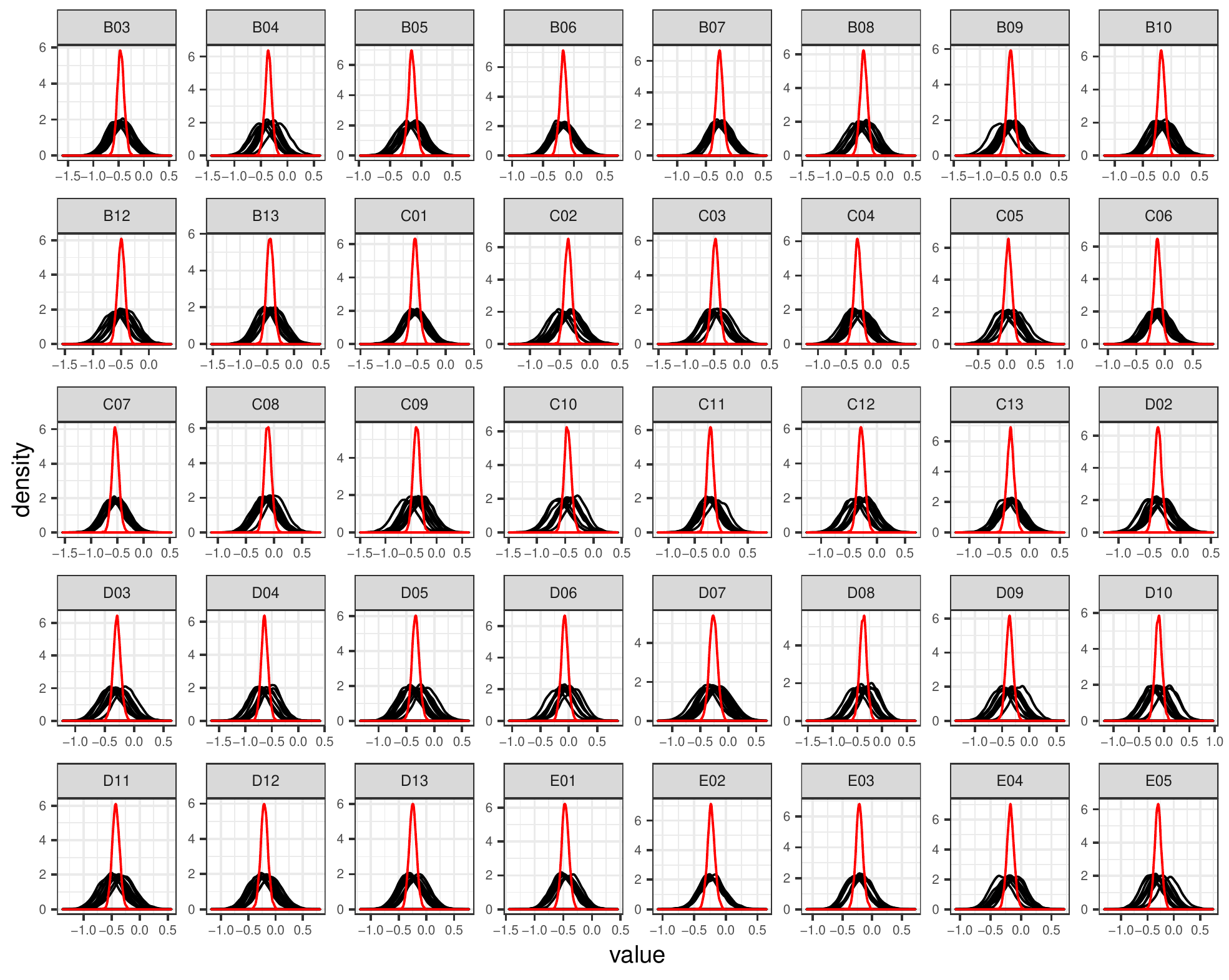}
     	\label{fig:site_densities_posterior}
\end{subfigure}

\begin{subfigure}[b]{1\textwidth}
\centering
\caption{Site difficulties vs the proportion of correct classification}
		\includegraphics[width=3.5in]{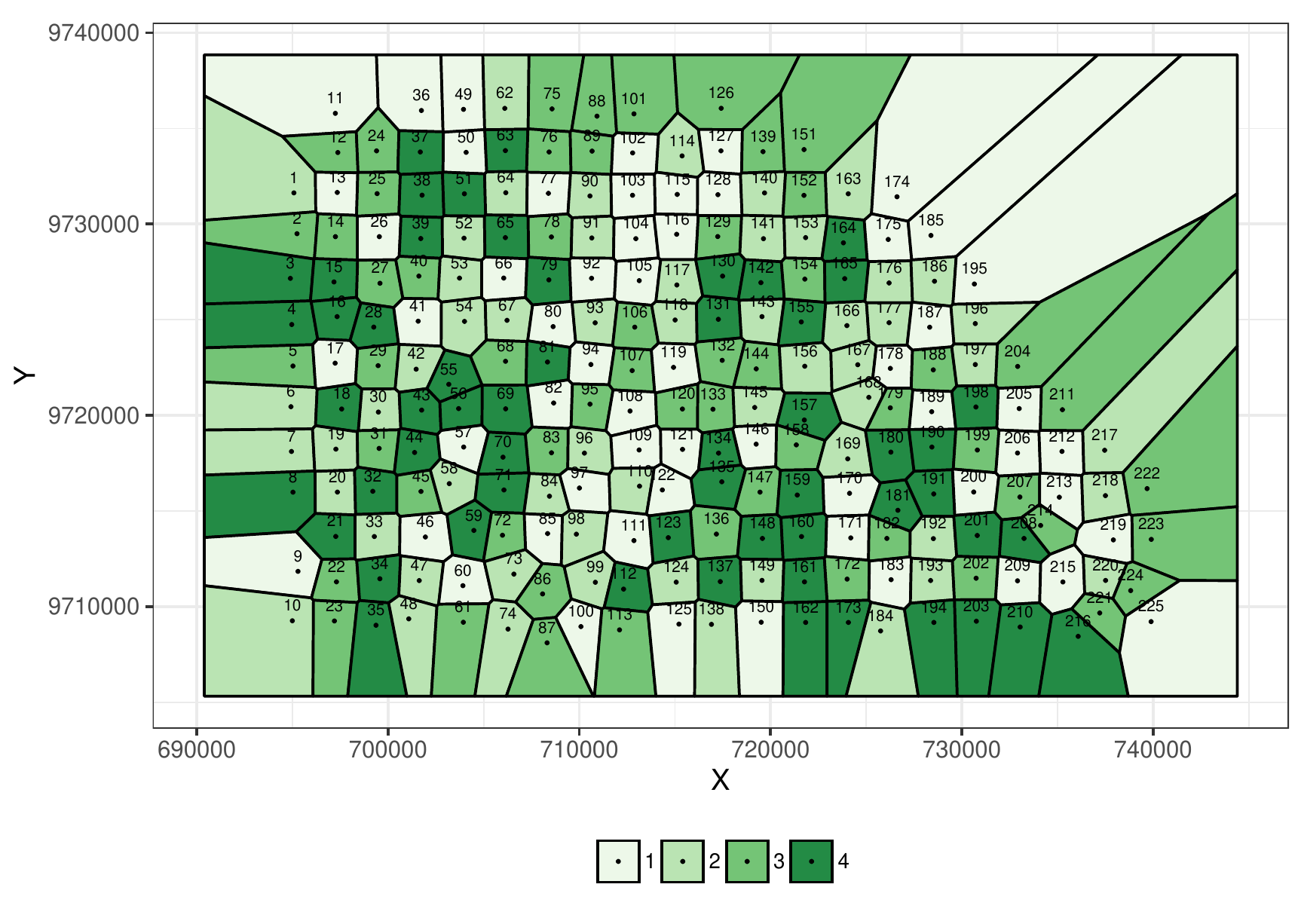}
   	\label{fig:voronoi_site_difficulties}
\end{subfigure}
	\caption{ {\scriptsize (a) Comparisons of the difficulties posterior estimates in the first 40 sites in the subsets and compared to the consensus estimate. 	(b) Voronoi diagram of the posterior estimates of the site-related difficulties. 
	Four categories were obtained using the quantiles where dark (light) green represent more difficult (easier) sites.
	Dots represent the location of the camera. } }
\end{figure}

Site-related difficulties were also produced. Images from some sites were harder to classify; see for instance C05 in Fig.\ref{fig:site_densities_posterior}.
A comparison between the 225 sites subposterior estimates among the 10 shards is presented in Fig.\ref{fig:site_ggpairs}. 
These results indicate a high degree of agreement between site difficulties obtained
from the different subsets and compared to the consensus values.  

In Fig \ref{fig:voronoi_site_difficulties} we show the posterior estimates of the site difficulties as a function of the proportion of correct classification and the Voronoi diagram of the site difficulties.
Sites with a larger difficulty estimate could indicate camera-related problems or installation issues. 


\section{Discussion and conclusion}

Networks with millions of contributors worldwide scattered in space and time engage in citizen science activities. 
This force is helping to reach the United Nations Sustainable Development Goals \citep{hsu2014development, fritz2019citizen}.
In the ecological context, citizens' contribution is two-fold: providing images, videos and audio files from animals, plants, etc.; but also producing classifications for estimating the distribution of species and for training machine learning algorithms. 
Yet, the data produced by citizens and hence by these programs are generally inaccurate and/or imprecise, which has received a lot of attention in recent years.
Prior research has also noted the importance of measuring the contribution of the subjects to citizen science projects \citep{silvertown2015crowdsourcing}. 
Although item-response models provide a natural framework in which to investigate and perhaps adjust for these challenges, 
the literature is limited on their usefulness for crowdsourced programs and particularly in the ecological context. 

In this research, we introduced a new methodology and family of IRT models specially tailored for ecological applications.
These methods can be used in a wide range of applications involving datasets elicited via crowdsourcing. 
We focus on ecological spatial data frequently collected in citizen science applications and specifically on citizen-elicited annotations of images.
We discuss the application of these models for real-world data, specifically in the presence of inconveniently large datasets.    
We found that our proposed framework of modeling outperforms the benchmark method in terms of precision and accuracy 
in a simulation experiment considering factors like the number of geotagged locations, the number of participants, etc.
Similarly, our approach performed better for real-life data. 
The application of the divide-and-recombine approach for massive item response problems is novel and can be
translated and extended to other settings.

Our methodology provides a better way of assessing the proficiency of the participants, 
which is useful for comparing them on equal grounds, 
for measuring their contribution, for building reputation schemes and for compensation purposes.
It also identifies challenging and commonly misclassified classes of responses (e.g. species) in order to produce better guidance, qualifications and users' training.   
Obtaining difficulties associated with images gives indications of cameras technical malfunctioning, misplacement,
light conditions, etc.  
We are currently developing the R package \textsf{staircase} (STAtistical Item Response models for Citizen Science applications in Ecology) that will allow practitioners to fit these new family of item response models, producing estimates of the participants' abilities along with other relevant ecological measures. 
The package adopts the Bayesian modeling framework and puts particular emphasis on big data.

In our models, participants' abilities were considered to be constant across the classification period.
This is reasonable since the period is generally short for many citizen science projects. 
However, in some circumstances this condition does not hold, e.g. an increased level of skill is achieved with time as participants learn and train.
Consequently, variations like those suggested in \citet{martin2002dynamic, dunson2003dynamic, wang2013bayesian} and \citet{weng2018real} can be adopted.
Furthermore, other methods accounting for the participants' reaction times could also be explored \citet[e.g. ][]{zhan2018cognitive, fox2010bayesian}.
Potential extensions of this particular case study based on the classification of camera traps images
include accounting for covariates affecting the species difficulties such as
the presence of young specimens or captured while moving.  
Our case study involves a relatively small geographical scale. 
Further work should consider items collected over broader scales surveys, at countries and continents levels, or even globally).

Camera-specific parameters and time of the day that also impact the quality of the photograph could be included as a factor in the model.
In terms of scalability, variational approximations could be used to speed up the computation on big datasets 
\citep{natesan2016bayesian, hui2017variational}.  
We are currently exploring an online updating approach for processing data as it becomes available \citep{schifano2016online, sarkka2013bayesian, callaghan2019optimizing}.

\section*{Acknowledgement}

This research was supported by the Centre of Excellence for Mathematical and Statistical Frontiers (ACEMS) and the Australian Research Council (ARC) Laureate Fellowship Program under the project ``Bayesian Learning for Decision Making in the Big Data Era'' (ID: FL150100150). 
The authors declare that they have no conflicts of interest. 
We thank Erin Peterson for the encouragement and the helpful discussions during the preparation of the manuscript. 
Data analysis and visualizations were undertaken in \textsf{R} software \citep{rprog} using the packages \textsf{rstan} \citep{rstan}, 
\textsf{tidyverse} \citep{tidyverse}, \textsf{ggvoronoi} \citep{ggvoronoi} and \textsf{ggrepel} \citep{ggrepel}.
\textsf{bayesplot} \citep{bayesplot}, 
and \textsf{ggforce} \citep{ggforce}.

\section*{Author Contributions}
ESF and KM conceptualized, designed the study and drafted the manuscript. 
ESF wrote the R codes. 
Both authors made a substantial contribution and approved the final submission.



\vspace{1cm}
\begin{center}
{\large\bf SUPPLEMENTARY MATERIAL} 
\end{center}
The supplementary material includes the following file:
\begin{description}
\item[Simulations results and posterior estimates obtained via consensus Monte Carlo.]
\end{description}

\begin{center}
{\large\bf Data and codes availability} 
\end{center}

The dataset and the R/Stan codes used in the case study (Section~ 4.3) can be found in the repository:  
\url{https://github.com/EdgarSantos-Fernandez/staircase}.
The following files are included:

\begin{description}
\item[serengeti.RDS]: contains the Serengeti data set.
\item[Item\_response\_modeling\_for\_big\_data.Rmd]: this R markdown file includes the full R/Stan codes for fitting the models, combining the posterior estimates and for visualizing the results.  
\item[Item\_response\_modeling\_for\_big\_data.pdf]: this is the knitted R markdown output file. 
\end{description}

\bibliography{ref}

\renewcommand\thefigure{S.\arabic{figure}}   
\renewcommand\thetable{S.\arabic{table}}
\renewcommand\thesection{S.\arabic{section}}

\newgeometry{inner=2.5cm,outer=2.5cm, top=2.5cm, bottom=2.5cm}

\section{SUPPLEMENTARY MATERIAL: Simulations results and posterior estimates obtained via consensus Monte Carlo}
\label{sim_res}

Summary: This is the supplementary material for Bayesian item response models for citizen science ecological data
and it is composed of two sections. 
Section \ref{sim_res} contains information supporting the simulation study. 
Some results regarding the Case Study are presented in Section \ref{posterior_consens}.

\begin{figure}[h]
\label{fig:abil_violin_3PL_3PLUS} 
\centering
\begin{subfigure}[h]{0.85\textwidth}
 \caption{3PL}
   \includegraphics[width=0.7\linewidth]{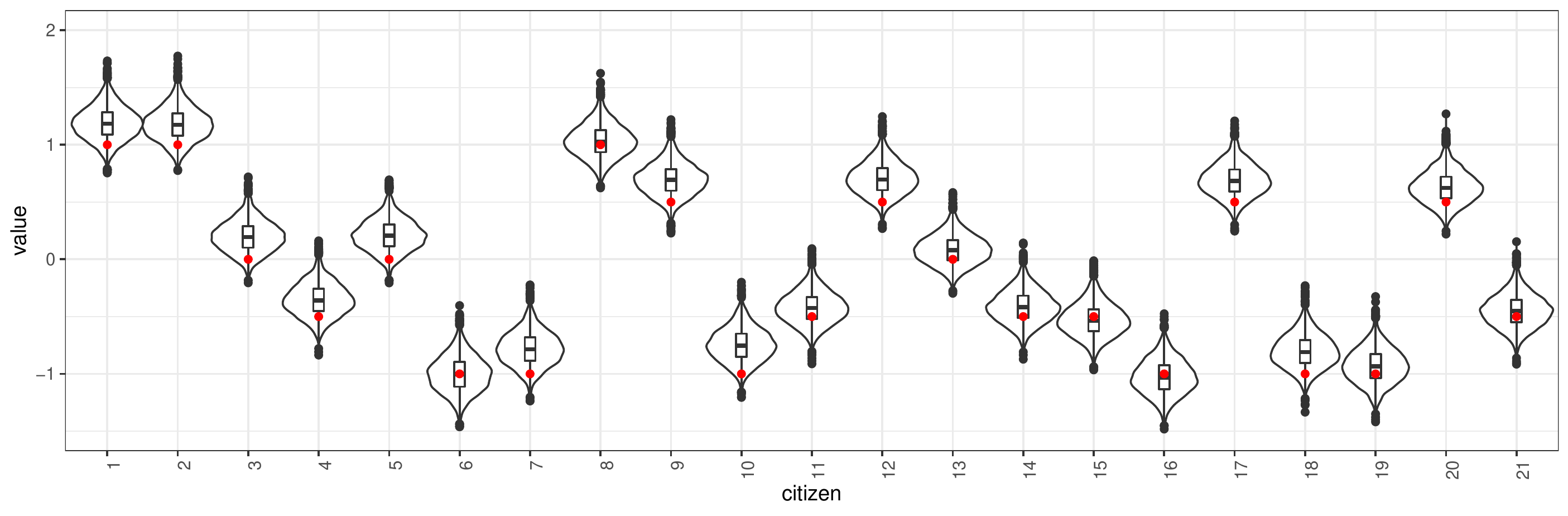}
    \label{fig:abil_violin_3PL} 
\end{subfigure}

\begin{subfigure}[h]{0.85\textwidth}
 \caption{3PLUS}
   \includegraphics[width=0.7\linewidth]{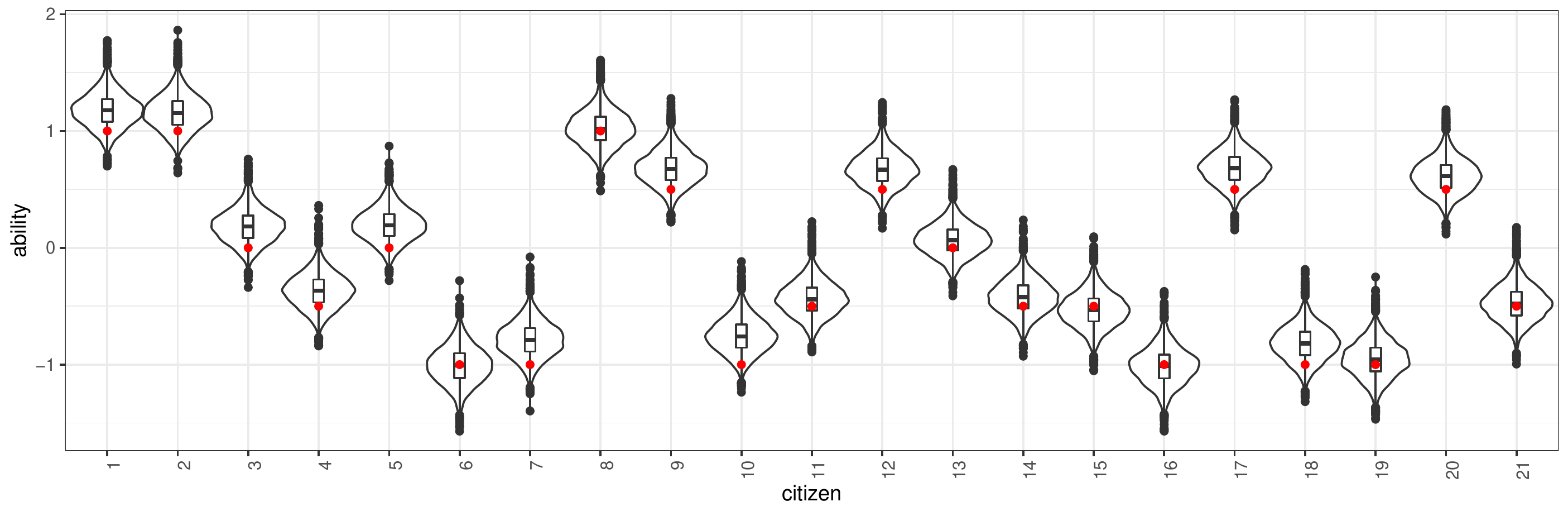}
   \label{fig:abil_violin_spat}
\end{subfigure}

\caption{Violin/box plot of the posterior distribution of the five groups of users' abilities $\theta_j$
in the 3PL (a) and 3PLUS model (b). The red dots represent the fixed true abilities. }
\end{figure}

Fig.\ref{fig:abil_violin_3PL} and \ref{fig:abil_violin_spat} depict the posterior distributions violin and boxplots for the abilities in both models (3PL vs 3PLUS) as a function of the true fixed value given by the red dots.

\begin{table}[ht]
\centering
\caption{Posterior summary statistics of the regression model for the $\log(\textrm{RMSE}_{\textrm{diff}})$.} 
\label{table:model_results}
\scalebox{0.70}{
\begin{tabular}{lrrrrrrrr}
Parameter & Rhat & n\_eff & mean & sd & se\_mean & 2.5\% & 50\% & 97.5\% \\ \hline
$b_\textrm{Intercept}$ & 1.0012 &  3614 & -0.2905 & 0.0786 & 0.0013 & -0.4452 & -0.2912 & -0.1376 \\ 
 $b_\textrm{model3PLUS}$ & 1.0006 &  3035 & -0.4437 & 0.0429 & 0.0008 & -0.5275 & -0.4439 & -0.3602 \\ 
 $b_\textrm{neli}$ & 1.0003 &  5022 & -0.0088 & 0.0027 & 0.0000 & -0.0141 & -0.0088 & -0.0035 \\ 
 $b_{min_{clas}}$ & 0.9999 &  3226 & -0.0021 & 0.0077 & 0.0001 & -0.0170 & -0.0022 & 0.0128 \\ 
 $b_{n_{cit}}$ & 0.9994 &  5967 & -0.0086 & 0.0010 & 0.0000 & -0.0106 & -0.0086 & -0.0066 \\ 
 $b_n$ & 1.0015 &  4111 & 0.0003 & 0.0001 & 0.0000 & 0.0001 & 0.0003 & 0.0005 \\ 
 $\sigma$ & 1.0010 &  2558 & 0.2774 & 0.0156 & 0.0003 & 0.2487 & 0.2766 & 0.3099 \\ 
  log-posterior & 1.0016 &  1590 & -28.1311 & 1.8649 & 0.0468 & -32.6632 & -27.8176 & -25.5008 \\ \hline
\end{tabular}
}
\end{table}

\newgeometry{inner=1.8cm,outer=1.8cm, top=1.8cm, bottom=1.8cm}
\begin{landscape}

\subsection{Simulations results.}
\label{sim_res2}

In section 3.2 we performed a simulation study to compare both models.

\begin{figure*}[htbp]
	\centering
		\includegraphics[width=8in]{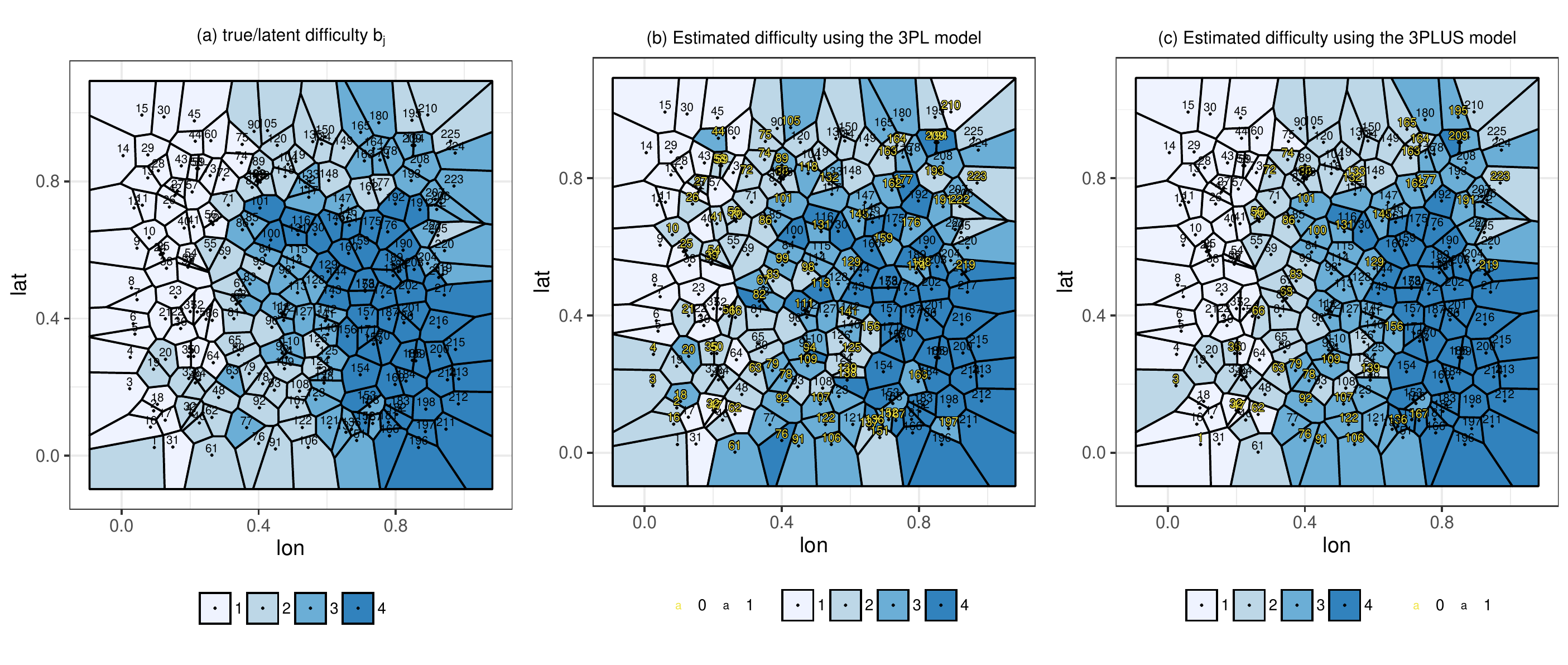}
	\caption{
	(a) Voronoi diagram of the latent spatially correlated item difficulties $b_j$. 
	The color represents the difficulty category given by the quantiles \{[-2.178, -0.514], (-0.514, 0.232], (0.232, 0.814], (0.814, 1.982]\} so that each category is composed of approximately 56 locations.
	The easiest and hardest groups are 1 and 4 respectively.
	(b) and (c) are the estimated item difficulties in the traditional 3PL and in the 3PLUS model. 
	The yellow labels represent incorrect category estimation.} %
	\label{fig:diff_voronoi}
\end{figure*}

\end{landscape}

\clearpage
\restoregeometry

\subsection{Posterior estimates obtained from the subposteriors via consensus Monte Carlo.}
\label{posterior_consens}

\begin{figure*}[h]
	\centering
		\includegraphics[width=3.5in]{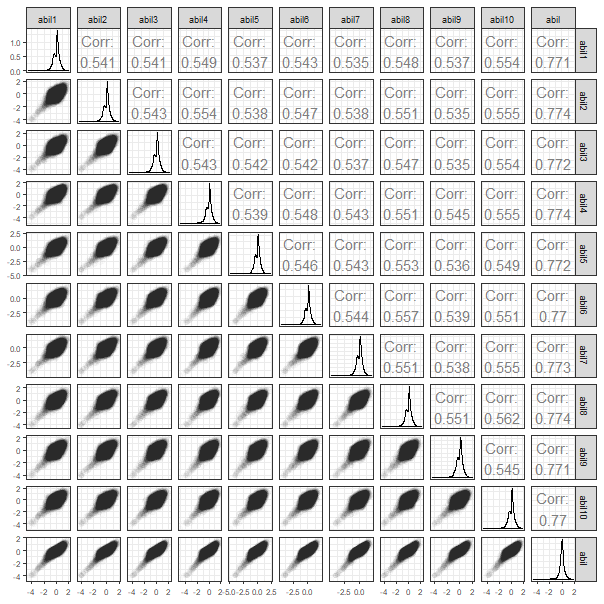}
	\caption{Comparison of the posterior estimates in the 21,347 users abilities in the shards and from the consensus approach.} 
	\label{fig:abil_ggpairs}
\end{figure*}

In the case study (section 4.3), we split the big dataset and fit the Bayesian item response model to each of the individual shards.
Fig \ref{fig:abil_ggpairs} shows a comparison of the ability estimates in the 10 shards and those obtained from the  
consensus Monte Carlo approach. 

\begin{figure*}[h]
	\centering
		\includegraphics[width=3.5in]{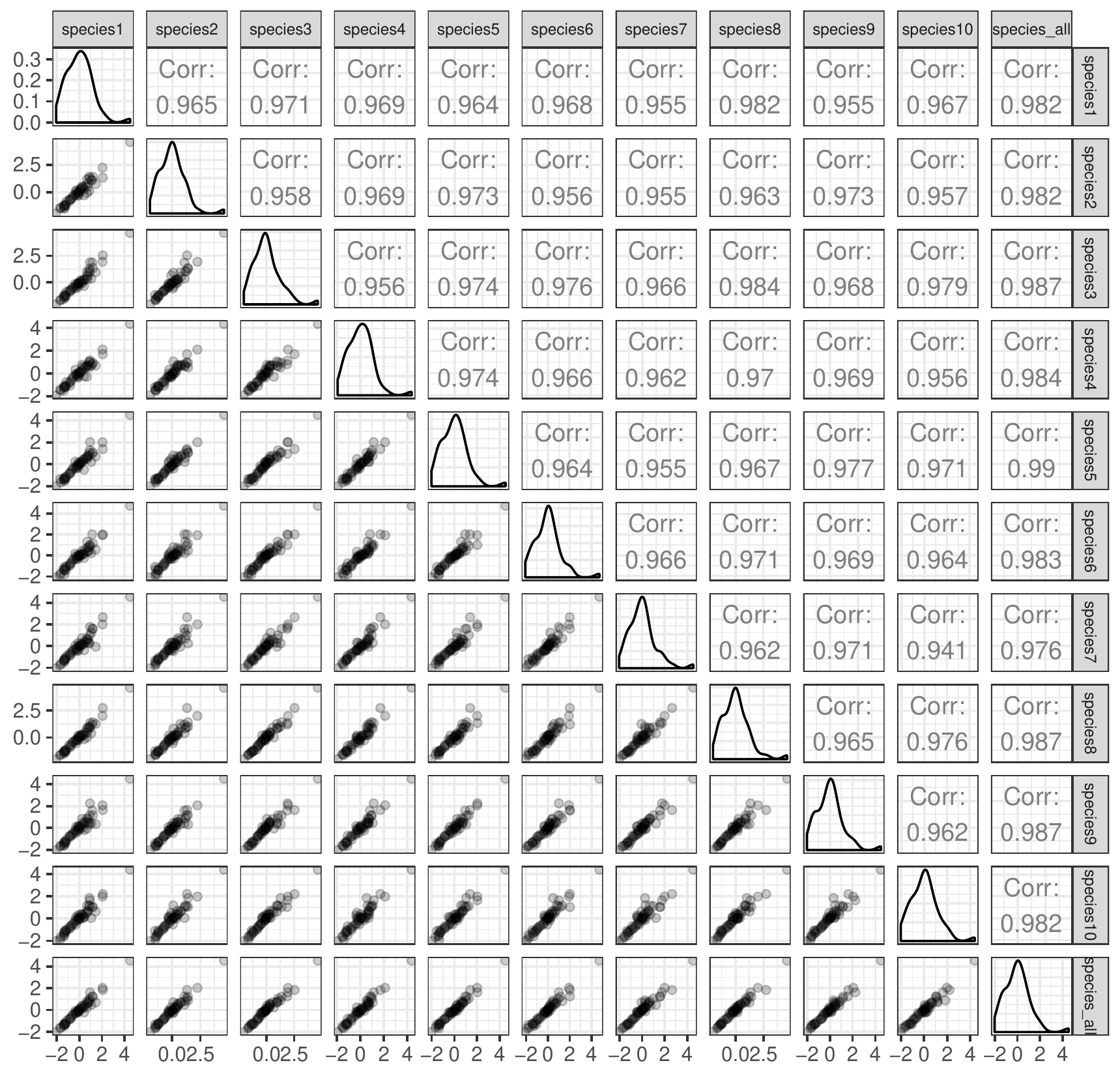}
	\caption{Comparisons of the posterior estimates of the species difficulties in the shards and from the consensus approach (species\_all).} 
	\label{fig:species_ggpairs}
\end{figure*}

\begin{figure*}[h]
	\centering
		\includegraphics[width=3.5in]{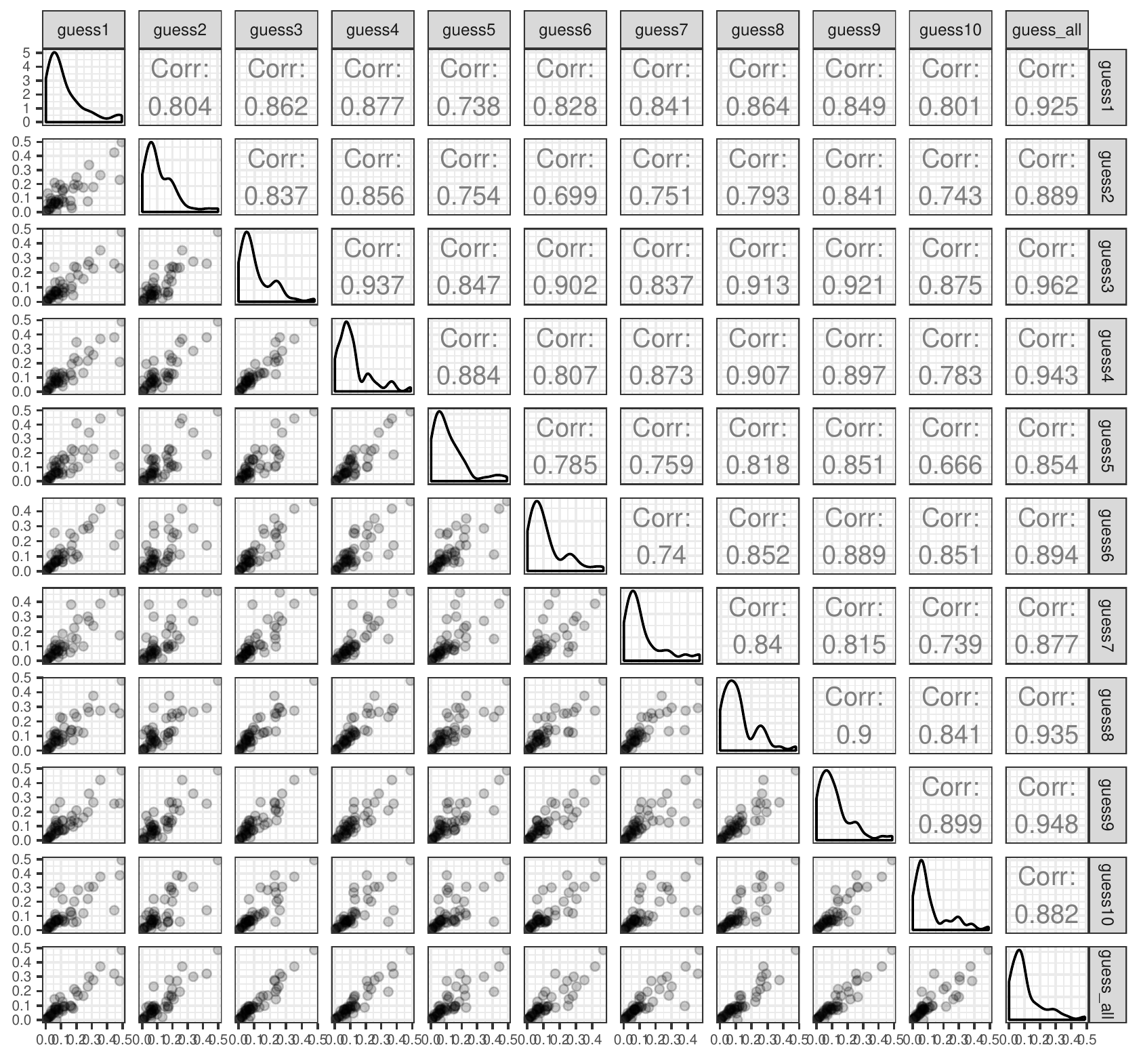}
	\caption{Comparisons of the posterior means estimates of the species pseudoguessing in the subsets and compared to the combined estimate (guess\_all). 
} %
	\label{fig:guess_species_ggpairs}
\end{figure*}

\begin{figure*}[h]
	\centering
		\includegraphics[width=3.5in]{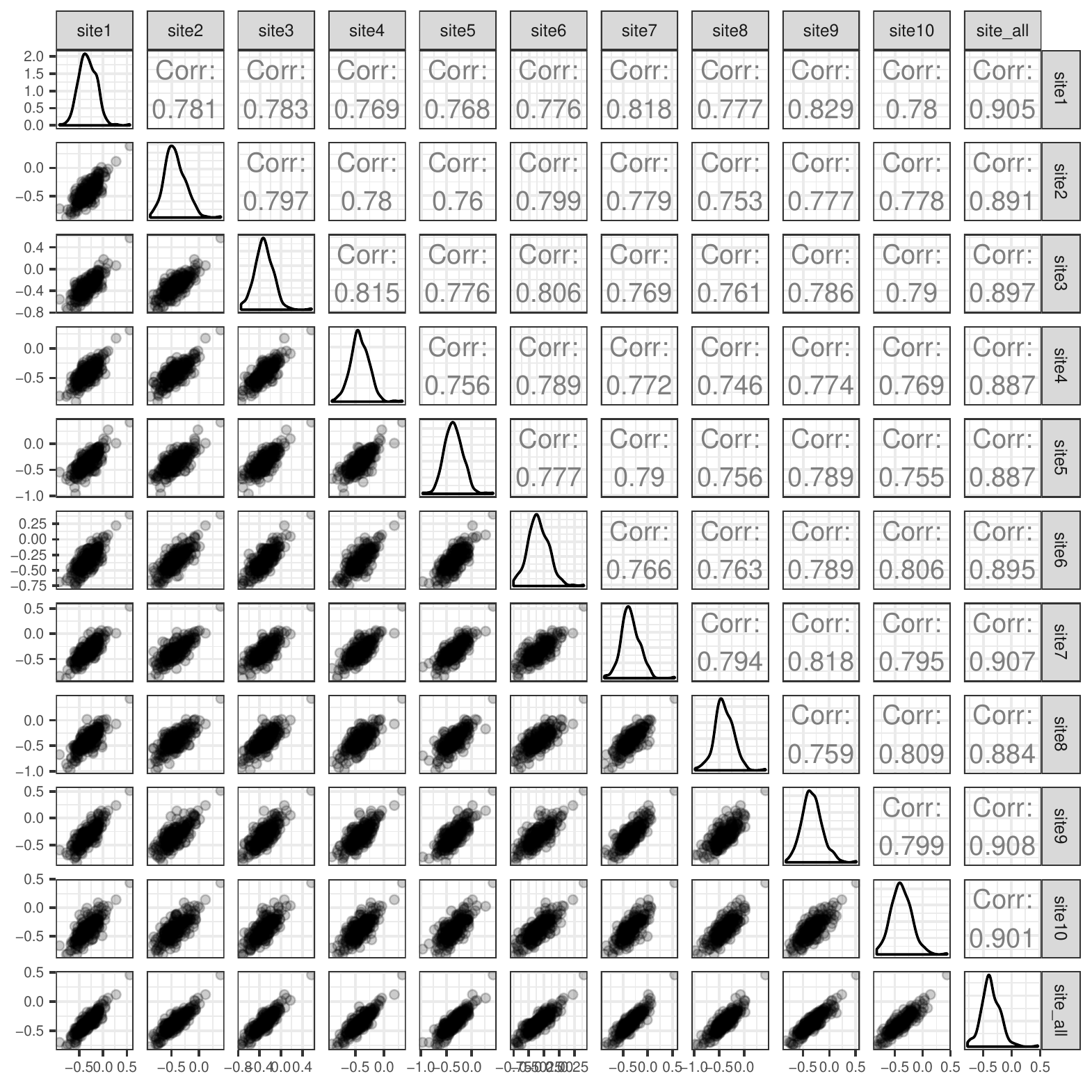}
	\caption{Comparisons of the difficulties posterior estimates in the 10 subsets (site1, site2, etc) and compared to the consensus estimate (site\_all). 
} %
	\label{fig:site_ggpairs}
\end{figure*}

Similar comparisons are shown in Fig \ref{fig:species_ggpairs}, \ref{fig:guess_species_ggpairs}  and \ref{fig:site_ggpairs}   
for the difficulties associated to the species, the pseudoguessing and the site difficulties respectively.

Table \ref{table:posterior_con} shows the consensus posterior estimates for some of the parameters of interest.

\begin{table}[h]
\caption{\label{table:posterior_con} Posterior estimates of some of the parameters via consensus Monte Carlo.}
\centering
\scalebox{0.60}{
\begin{tabular}{rllrrrrr}
  \hline
 & param & id & mean & sd & med & $q_{2.5}$ & $q_{97.5}$ \\ 
  \hline
1 & diff\_species\_3 & baboon & -0.802 & 0.229 & -0.807 & -1.251 & -0.331 \\ 
  2 & diff\_species\_24 & hyenaStriped & 2.007 & 0.734 & 1.902 & 0.842 & 3.697 \\ 
  3 & diff\_species\_26 & impossible & 4.535 & 0.614 & 4.495 & 3.433 & 5.898 \\ 
  4 & diff\_species\_30 & lionFemale & -0.046 & 0.267 & -0.046 & -0.585 & 0.482 \\ 
  5 & diff\_species\_49 & zebra & -2.020 & 0.180 & -2.019 & -2.390 & -1.671 \\ 
  6 & diff\_site\_1 & B03 & -0.456 & 0.214 & -0.456 & -0.882 & -0.023 \\ 
  7 & diff\_site\_2 & B04 & -0.359 & 0.216 & -0.360 & -0.788 & 0.085 \\ 
  8 & diff\_site\_3 & B05 & -0.151 & 0.190 & -0.150 & -0.527 & 0.237 \\ 
  9 & diff\_site\_4 & B06 & -0.167 & 0.183 & -0.169 & -0.526 & 0.208 \\ 
  10 & pseudoguessing\_3 & baboon & 0.137 & 0.083 & 0.127 & 0.009 & 0.322 \\ 
  11 & pseudoguessing\_24 & hyenaStriped & 0.230 & 0.100 & 0.241 & 0.019 & 0.397 \\ 
  12 & pseudoguessing\_26 & impossible & 0.010 & 0.009 & 0.007 & 0.000 & 0.034 \\ 
  13 & pseudoguessing\_30 & lionFemale & 0.366 & 0.074 & 0.373 & 0.203 & 0.497 \\ 
  14 & pseudoguessing\_49 & zebra & 0.009 & 0.009 & 0.007 & 0.000 & 0.032 \\ 
  15 & abil\_1 & 1ee963bb67658ec7fb7ff5568f2e5c52 & 1.016 & 0.587 & 0.981 & -0.052 & 2.288 \\ 
  16 & abil\_2 & 43047c40867c00a83d0a3982f12f515b & -0.446 & 0.822 & -0.443 & -2.128 & 1.191 \\ 
  17 & abil\_3 & 52a791d0b1715d3a64e3304853d78c7c & 1.019 & 0.386 & 0.978 & 0.353 & 1.893 \\ 
  18 & abil\_4 & 6ed2e808d62f729c45912856bffb1ae7 & 0.000 & 0.592 & -0.040 & -1.067 & 1.315 \\ 
  19 & abil\_5 & 80c3e443a01284ce8be767b3b68b921d & 1.176 & 0.354 & 1.143 & 0.555 & 1.958 \\ 
  20 & abil\_6 & b1132cb9f684ac9bf8b08522c7537ea6 & -0.041 & 0.528 & -0.069 & -1.020 & 1.122 \\ 
  21 & abil\_7 & cb7671ced14ad6f669a2601b249e0e93 & -1.407 & 0.580 & -1.402 & -2.594 & -0.252 \\ 
  22 & abil\_8 & cb93d3701ea002dd736828eb1a87cb88 & -0.880 & 0.616 & -0.884 & -2.107 & 0.363 \\ 
  23 & abil\_9 & e6b9208fb9d03ce1c3885fbb866a39f8 & 0.226 & 0.458 & 0.200 & -0.620 & 1.200 \\ 
   \hline
\end{tabular}
}
\end{table}

\end{document}